\documentclass[twocolumn, tighten]{aastex701}

\usepackage{amsmath}
\usepackage{courier}
\usepackage{listings}
\usepackage{gensymb}
\usepackage{bm}
\usepackage{physics}
\usepackage{subfigure}
\usepackage{subcaption}
\usepackage{xspace}

\def\m87{M87$^*$\xspace}
\def\sgra{Sgr~A$^*$\xspace}

\defcitealias{EHT2022-I}{Sgr~A*~I}
\defcitealias{EHT2022-II}{Sgr~A*II}
\defcitealias{EHT2022-III}{Sgr~A*~III}
\defcitealias{EHT2022-IV}{Sgr~A*~IV}
\defcitealias{EHT2022-V}{Sgr~A*~V}
\defcitealias{EHT2022-VI}{Sgr~A*~VI}
\defcitealias{EHT2024-VII}{Sgr~A*~VII}
\defcitealias{EHT2024-VIII}{Sgr~A*~VIII}

\defcitealias{EHT2019-I}{\m87~I}
\defcitealias{EHT2019-II}{\m87~II}
\defcitealias{EHT2019-III}{\m87~III}
\defcitealias{EHT2019-IV}{\m87~IV}
\defcitealias{EHT2019-V}{\m87~V}
\defcitealias{EHT2019-VI}{\m87~VI}

\begin{document}

\title{Measuring the Black Hole and Accretion Parameters of Sagittarius\,A$^*$ from EHT Observations using a Semi-Analytic Model}


\correspondingauthor{Braden J. Marazzo-Nowicki}
\author[orcid=0009-0001-8191-4538,gname=Braden, sname='Marazzo-Nowicki']{Braden J. Marazzo-Nowicki}
\affiliation{Department of Astronomy, University of Maryland, College Park, MD 20742, USA}
\affiliation{Center for Astrophysics $|$ Harvard \& Smithsonian, 60 Garden Street, Cambridge, MA 02138, USA}
\affiliation{Black Hole Initiative at Harvard University, 20 Garden Street, Cambridge, MA 02138, USA}
\email[show]{bnowicki@terpmail.umd.edu}

\author[orcid=0000-0003-3826-5648, gname=Paul, sname='Tiede']{Paul Tiede}
\affiliation{Center for Astrophysics $|$ Harvard \& Smithsonian, 60 Garden Street, Cambridge, MA 02138, USA}
\affiliation{Black Hole Initiative at Harvard University, 20 Garden Street, Cambridge, MA 02138, USA}
\email[]{ptiede@fas.harvard.edu}  

\author[orcid=0000-0001-9939-5257,gname=Dominic O., sname='Chang']{Dominic O. Chang}
\affiliation{Center for Astrophysics $|$ Harvard \& Smithsonian, 60 Garden Street, Cambridge, MA 02138, USA}
\affiliation{Black Hole Initiative at Harvard University, 20 Garden Street, Cambridge, MA 02138, USA}
\affiliation{Department of Physics, Harvard University, Cambridge, Massachusetts 02138, USA}
\email[]{dochang@g.harvard.edu} 

\author[orcid=0000-0002-7179-3816,gname=Daniel C. M., sname='Palumbo']{Daniel C. M. Palumbo}
\affiliation{Center for Astrophysics $|$ Harvard \& Smithsonian, 60 Garden Street, Cambridge, MA 02138, USA}
\affiliation{Black Hole Initiative at Harvard University, 20 Garden Street, Cambridge, MA 02138, USA}
\email[]{daniel.palumbo@cfa.harvard.edu} 

\author[0000-0002-4120-3029]{Michael D. Johnson}
\affiliation{Center for Astrophysics $|$ Harvard \& Smithsonian, 60 Garden Street, Cambridge, MA 02138, USA}
\affiliation{Black Hole Initiative at Harvard University, 20 Garden Street, Cambridge, MA 02138, USA}
\email[]{mjohnson@cfa.harvard.edu} 

\begin{abstract}


The Event Horizon Telescope (EHT) Collaboration produced the first image of the apparent shadow of the central black hole of Sagittarius\,A$^*$ (\sgra). \sgra source structure varies significantly on timescales shorter than the duration of an observation, preventing improved data coverage through Earth rotation aperture synthesis. This rapid variability provides the opportunity to quantify intrinsic variability and separate time-variable emission features from stable signatures of strong gravity and the accretion environment. To infer the properties \sgra and its surrounding accretion flow, we perform Bayesian inference on a series of EHT data segments (``snapshots''). We directly fit parameters of a semi-analytic emission model jointly with complex station gains to snapshot visibilities, then extract estimates of the time-averaged, persistent source structure and temporal variability by stacking snapshots in a Bayesian hierarchical model. This approach successfully reproduces parameters of General Relativistic Magnetohydrodynamics simulations using synthetic EHT observations. Even with physically motivated assumptions about the \sgra environment, black hole spin and magnetic field parameters are poorly constrained by 2017 EHT observations. Our inference constrains other parameters, favoring a nearly face-on observer inclination ($\theta_{\rm o} = 9.2\degree \pm 3.6 \degree \pm_{\rm v} 11.6\degree$), an emission peak near the horizon ($R_{\rm peak} = 4.9 \pm 0.1 \pm_{\rm v} 0.5\,GM/c^2$), near-vertical projected spin position angle ($p.a. = 7.3\degree \pm 7.08 \degree \pm_{\rm v} 43.5\degree$ counterclockwise from vertical),
and dominant emission $43.4\degree \pm 2.0\degree \pm_{\rm v} 5.9\degree$ above the equatorial plane, where we separate average structure uncertainty ($\pm$) from the impacts of temporal variability and model misspecification ($\pm_{\rm v}$).


\end{abstract}



\section{Introduction} \label{sec: introduction}
In 2017, the Event Horizon Telescope (EHT) used an 8-element Very Long Baseline Interferometry (VLBI) array operating at $\lambda = 1.3$\,mm to observe Sagittarius A$^*$ (\sgra), the radio source associated with the Supermassive Black Hole (SMBH) at the center of the Milky Way \citep{1974ApJ...194..265B}. Image reconstructions displayed a bright ring of emission around a central brightness depression, or apparent ``shadow''---features consistent with expectations for plasma accretion flows in a Kerr spacetime \citep[][hereafter Sgr~A* I-VIII]{EHT2022-I, EHT2022-II, EHT2022-III, EHT2022-IV, EHT2022-V, EHT2022-VI, EHT2024-VII, EHT2024-VIII}.

In particular, this image morphology with a distinct ``shadow'' is a general prediction of models of optically thin plasmas around SMBHs as viewed by a distant observer 
\citep[e.g.][]{1973blho.conf..215B, 1979A&A....75..228L, 2000CQGra..17..123D, 2000ApJ...528L..13F, Broderick_2016, 2019ApJ...885L..33N}.
In previous EHT studies, analysis in the visibility domain characterized morphological parameters of the likely ring-like image (e.g.\ diameter, orientation, thickness); in these analyses, the ring width was well constrained but the amplitude and orientation of ring asymmetry were poorly constrained and dependent on measurement methods \citepalias{EHT2022-IV}. Comparison with General Relativistic Magnetohydrodynamics (GRMHD) simulations led to strong constraints on \sgra magnetization but weak constraints on black hole spin \citepalias{EHT2022-V}. Beyond these broad conclusions, the physical conditions of the accretion flow and near-horizon spacetime structure of the \sgra emitting region are largely unconstrained.
 
\sgra ($M \approx 4 \times 10^6 M_\odot$) 
has a typical gravitational timescale $GM/c^3 \approx 20$s and time variability on the order of minutes to hours \citep[][see also \citetalias{EHT2019-I, EHT2022-IV}]{2019Sci...365..664D, 2019A&A...625L..10G, 2020A&A...636L...5G}. Traditional VLBI imaging relies on Earth-rotation aperture synthesis to increase sampling of the ($u, v$) plane: as Earth rotates, baselines change their orientation with respect to the source, enabling greater Fourier coverage \citep{2017isra.book.....T}. When source structure evolves on shorter timescales than a several-hour EHT observation, this technique cannot be employed to increase sampling of the ($u, v$) plane and accurately reproduce the observed data. For \sgra, the EHT found the magnitude of variability to be a substantial fraction of the correlated flux density in some cases \citepalias{EHT2022-IV}. This rapid intraday variability complicates attempts to infer physical parameters: visibility data is sparse and time variability must be handled in an informed manner. One approach to this problem---``snapshot modeling''---involves fitting parameters of interest over short snapshot observations of a source structure which is expected to be static.

\begin{figure*}[!htb]
    \centering
    {\includegraphics[width=0.95\textwidth]{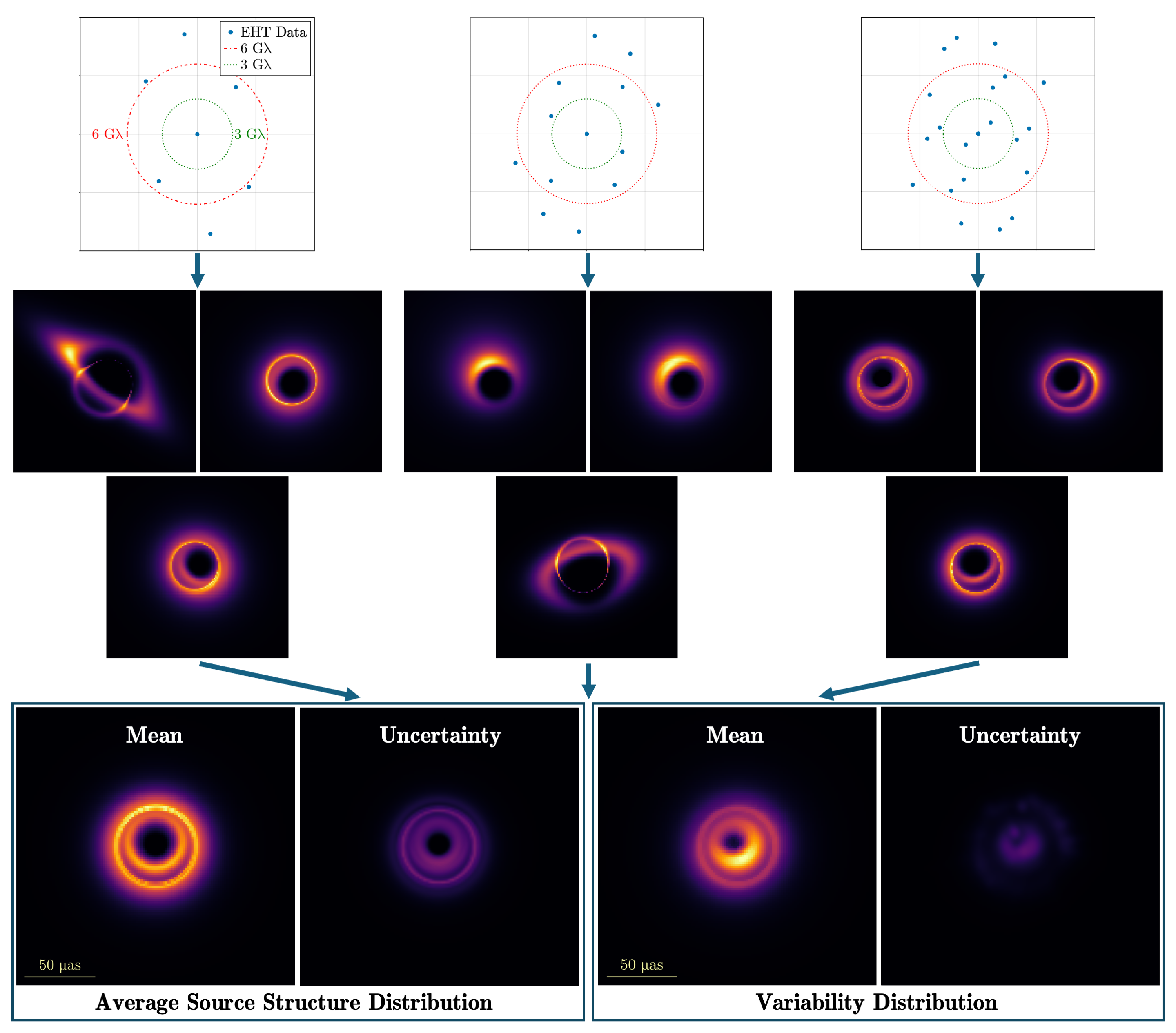}} 
    \caption{Schematic depiction of the snapshot modeling methodology. We perform inference on complex visibilities from individual snapshots of varying coverage, jointly inferring physical model parameters and complex station gains. The resulting posteriors are then stacked in a global Bayesian hierarchical model, giving estimates for the average source structure and variable features. Top row, left to right: example ($u,v$)-coverage for individual snapshots, with snapshots ranging between poor coverage (a minimum of four baselines), such as the snapshot indexed as scan 27 (April 7, 2017 at 06:03); medium coverage, such as scan 114 (April 7, 2017 at 11:23); and good coverage (a maximum of 9 baselines), such as scan 157 (April 7, 2017 at 13:49 UTC). Even for the best-sampled snapshots, coverage is very sparse. Data taken at $227.07$ GHz are shown in blue and baselines of 3 G$\lambda$ and 6 G$\lambda$ are shown in green and red, respectively. Middle row, left to right: images drawn from the inferred posterior distributions corresponding to snapshot visibility data (scan 27, scan 114, and scan 157, respectively). Due to the sparsity of data in a single snapshot, many parameters are unconstrained and myriad different image morphologies fit the data. These morphologies contribute to structures in the mean and standard deviation images. Bottom row, left to right: mean and standard deviation (uncertainty) images from the global posterior, a Bayesian hierarchical model stacking the posteriors from every snapshot, with an average source structure distribution and a variability distribution. The mean image from the average source structure distribution indicates persistent morphological features across snapshots. The mean variability map shows the image features most subject to intra-snapshot variability.}
    \label{fig:method-schematic}
\end{figure*}

Snapshot modeling of \sgra using the EHT observations was previously performed using a series of simple geometric models, finding that models with a ring-like structure best fit the emission from \sgra \citepalias{EHT2022-IV}. The ring diameter was well constrained, but other morphological parameters---like brightness asymmetries---were measured with less confidence \citepalias{EHT2022-IV}. Beyond inferring morphological parameters of on-sky emission, directly probing parameters of the physical accretion environment is of great interest for further understanding of the source. In this paper, we employ a lightweight, robust, phenomenological dual-cone emission model which provides a suitable and quick approximation to images generated from GRMHD simulations \citep{Chang2024}. We use parallel tempered Markov Chain Monte Carlo (PTMCMC) sampling to fit the dual-cone model to visibility data, exploring the posteriors of physics-based model parameters. We then stack the snapshot posteriors in a Bayesian hierarchical model to form a global posterior. \autoref{fig:method-schematic} depicts the snapshot modeling scheme, whereby inference on individual snapshot visibilities produces parameter posteriors that inform a global hierarchical model.

In Section \ref{sec:model}, we discuss the semi-analytic emission model and its purpose in greater detail. Observations are briefly discussed in Section \ref{sec:observations}. In Section \ref{sec:inference}, we consider methods of fitting EHT observations of \sgra and stacking posteriors across snapshots. In Section \ref{sec:results}, we outline results for a single snapshot, then the hierarchical model formed from all snapshots, and finally inference on synthetic data where the ground truths of several parameters are known. In Section \ref{sec:discussion}, we consider the limitations of this analysis and the scientific implications of the results. Finally, we summarize our results in Section \ref{sec:conclusions}.

\section{Semi-Analytic Accretion \& Emission Model} \label{sec:model}

GRMHD simulations are essential for interpreting observations of Low-Luminosity Active Galactic Nuclei (LLAGN), including the EHT observations of \sgra and M87$^*$. Beyond GRMHD simulations, there is a rich history of simulations exploring synchrotron-emitting plasma flows and accretion near the galactic center \citep[][]{Yuan_2004, Broderick_2009,Dexter_2009,Broderick_2011,Pu_2016, ipole, EHT2022-V, Galishnikova_2023, Chan:2024bjg}. The computational demands of GRMHD simulations make them unsuitable for directly fitting observational data, whereby sophisticated sampling is performed in complicated, high-dimensional posteriors. 
This difficulty motivates an ongoing effort to develop lightweight semi-analytic emission models \citep[some examples include][]{Broderick_2016, Narayan2021,Papoutsis_2023} or neural surrogate models for GRMHD \citep[][]{Farah_2025, Franc} that can be used to describe and infer properties from these systems. 
Semi-analytic fits for \sgra have been previously performed using proto-EHT observations \citep[see, e.g.][]{Broderick_2016}, but these did not account for intraday source variability. 

In this paper, we use the dual-cone emission model in \citealt{Chang2024}, which expands on a similar equatorial plane emission model in \citealt{Palumbo2022}. The dual-cone model considers synchrotron-emitting, magnetized plasma constrained to flow within two thin, oppositely oriented cones aligned with the black hole's spin axis. 
The model is inherently axisymmetric. The cones open at an angle $\theta_{\rm s}$ from the spin axis. Emission is ray-traced in a Kerr geometry to a distant observer at inclination angle $\theta_{\rm o}$ from the spin axis. Several black hole, magnetic field configuration, and fluid configuration parameters inform the model. The model sufficiently reproduces images generated from a wide variety of GRMHD simulations while remaining efficient enough for Bayesian inference of model parameters on visibility data \citep{Chang2024}. We therefore perform a kind of \textit{photogrammetry}---reconstructing the three-dimensional physical structure and environmental conditions of \sgra from EHT data. 

The spacetime structure in the model is calculated using the Kerr metric for a black hole of dimensionless spin $a_{\rm *}$. 
The mass-to-distance ratio, $\theta_{\rm g}$ specifies a stretch modification on the image. 
A projected position angle $p.a.$ relative to the spin axis can be specified. 
Spin position angle $p.a.$ is measured East of North, or counterclockwise of vertical. As discussed, $\theta_{\rm o}$ specifies the inclination angle of a distant observer relative to the black hole spin axis and $\theta_{\rm s}$ specifies the opening angle of the cones relative to the spin axis. Unlike some previous EHT prescriptions, where positive spin indicates prograde fluid flow and negative spin retrograde, the sign of the spin in this model indicates whether it is aligned or anti-aligned with the observer's line of sight.

The plasma flow is specified by the fluid speed in a Zero Angular Momentum Observer (ZAMO) frame, $\beta_{\rm v}$, and the fluid velocity azimuthal angle in the ZAMO frame, $\chi$. 
Magnetic field lines of the magnetic field $\mathbf{B}$ thread the emission model. The magnetic field configuration is specified with $\iota$, the angle of field lines with respect to the orthogonal of emission planes in the fluid frame, and $\eta$, the azimuthal-tangential angle in the fluid frame. We prescribe a power law synchrotron emissivity profile with $R_{\rm peak}$, the characteristic radius of emission, and $p_1$ and $p_2$, the inner and outer exponents of the emissivity profile. 
The spectral index of emission is $\sigma$. 
Emission is parallel transported along null geodesics to the distant observer.  
For this analysis, we use the resulting Stokes I intensity images containing both the $n=0$ (direct emission) and $n=1$ (a component of the indirect emission comprising the photon ring; see \citealt{Johnson2020}) sub images, though other Stokes parameter maps incorporating polarization may also be produced from the model. Images produced with the dual-cone model typically feature a distinct brightness depression, or ``shadow,'' diffuse direct emission, and a thin, bright photon ring structure \citep[for more discussion of the black hole photon ring, see, e.g.][]{Johnson2020, PhysRevD.102.044017, 2019PhRvD.100b4018G}.

Though General Relativity predicts a direct relationship between the spin and mass-to-distance ratio---the angular gravitational radius, $\theta_g \equiv GM/(c^2D)$---of a black hole and its asymptotic shadow diameter and shape \citep{1973blho.conf..215B}, there are numerous degeneracies between black hole parameters and those of the emitting plasma for the direct emission \citep[see, e.g.][]{EHT2019-V,EHT2019-VI, 2019PhRvD.100b4018G}. When attempting to constrain black hole parameters, it is therefore advantageous to perform Bayesian inference over a large parameter space and model the black hole and its emission parameters simultaneously. By exploring posteriors for many parameters, we may accurately characterize parameter values and uncertainties and better understand where degeneracies arise.

The semi-analytic model is ultimately an approximation of the physical environment around an SMBH and, therefore, has several key limitations. Conical emission surfaces are infinitely thin, and no optical depth prescription or full radiative transfer scheme is employed. While this enables fast image production and Bayesian inference on VLBI data, it limits the applicability of the model to relatively optically thin sources at \textit{mm} wavelengths, misspecifying potentially optically thick sources. Time-variable structures/turbulence and asymmetries are not modeled. The model captures time-averaged structures and axisymmetric flows and is less flexible for analysis of time-variable structures in individual snapshots. The model uses several other simplifying assumptions: the emission region is compact; there is no radial evolution of the plasma velocity in the ZAMO frame, $\beta_{\rm v}$; and, as previously described, the emissivity model is a simplified, radially dependent double power law defined by $R_{\rm peak}$, $p_1$, and $p_2$ \citep{Chang2024}. Snapshot inference on GRMHD simulations and \sgra data are subject to biases and systematic errors due to the lack of structural turbulence in the model \citep{Chang2024}. Real data also include degrees of freedom not present in the dual-cone model. Nonetheless, the model's ability to capture time-averaged GRMHD source structure motivates its use in this analysis. For further discussion about the dual-cone emission model, see \citealt{Chang2024} and Appendix \ref{sec:model-appendix}.

\subsection{Image Degeneracies} \label{subsubsec:degeneracies}

There are degeneracies in most image features---certain alterations could arise from a combination of a variety of physical parameters. For instance, the mass-to-distance ratio, characteristic emission radius, and cone opening angle can all resize the resulting image (the cone opening angle has a greater effect on the direct emission and shadow). The spin, fluid and magnetic field parameters ($\chi$, $p_1$, $p_2$, $\iota$, $\beta_{\rm v}$), and the inclination strongly impact image morphology and brightness asymmetries. For further discussion of the semi-analytic emission model and the impacts of parameters on resulting images, see Appendix \ref{sec:model-appendix}. In this analysis, we fix the mass-to-distance ratio, eliminating, for instance, the degeneracy between mass-to-distance ratio and characteristic emission radius. However, other degeneracies remain, leading to complicated, multimodal posteriors in several variables---especially with sparse data in each snapshot. It is therefore important to accurately and thoroughly explore the posteriors of each parameter so we characterize all modes of interest.

\section{Observations and Data Products} \label{sec:observations}

\subsection{VLBI Data Products} \label{subsec:dataprods}

VLBI observations do not directly sample sky brightness distributions. Instead, at a time $t$, they sample the complex visibilities of a source image $I(\boldsymbol{x},t)$---the Fourier transform of the image---as originally described by the van Cittert-Zernike theorem:
\begin{equation}
    \tilde{I}(\boldsymbol{u}, t) = \iint e^{2\pi i \boldsymbol{u} \cdot \boldsymbol{x}} I(\boldsymbol{x},t) \dd[2]{\boldsymbol{x}}.
\end{equation}
Here, $\boldsymbol{x} = (x,y)$ are angular coordinates on the sky image and $\boldsymbol{u} = (u,v)$ are coordinates projected along baselines in units of the observing wavelength \citep{2017isra.book.....T}. For a given baseline with coordinate $\bm{u}_{ab}$, we denote the ideal visibility by $\tilde{I}_{ab}$. 

However, the observed visibilities $\hat{V}_{ab}$ are affected by factors introducing statistical and systematic errors, such as refraction and disturbances in the local atmosphere, instrumental bias, and cable length delay at individual telescope sites. We model these alterations in observed amplitude and phase as complex station gains $g$ at each specific telescope ($g_a$ and $g_b$ here) for each timestamp. The observed visibility $\hat{V}_{ab}$ correlated on a baseline between stations $a$ and $b$ is given by
\begin{equation}\label{eq:vis_obs}
    \hat{V}_{ab} = g_ag_b^*\tilde{I}_{ab} + \epsilon_{ab} = \mathcal{V}_{ab} + \epsilon_{ab},
\end{equation}
where $\mathcal{V}_{ab} = g_a g_b^* \tilde{I}_{ab}$ is the model corrupted visibility and $\epsilon_{ab}$ is the statistical error on the baseline. We assume that $\epsilon_{ab}$ is a zero-mean, circularly symmetric, complex Gaussian random variable:
\begin{equation}\label{eq:vis_noise}
    \epsilon_{ab} \sim \mathcal{N}(0,\sigma_{ab}^2)+i\mathcal{N}(0,\sigma_{ab}^2),
\end{equation}
where $\sigma_{ab}$ is the statistical uncertainty for the baseline $ab$. 



\subsection{EHT Observations of \sgra} \label{subsec:obs-SgrA}

For this analysis, we used EHT observations of \sgra taken on April 7, 2017---the day with the greatest $(u,v)$-coverage in the 2017 observational campaign \citepalias{EHT2022-III}. The data were preprocessed in a manner similar to the preprocessing pipeline in \citetalias{EHT2022-III} and lightcurve normalized using the Atacama Large Millimeter/Submillimeter
Array (ALMA) \sgra lightcurve data \citepalias{EHT2022-IV}. In addition, to account for residual calibration uncertainties, refractive scattering in the interstellar medium, phase variations across the measurement frequency band, and leakage between right- and left-circularly polarized data, we inflate the statistical error budget $\sigma_{sb}$ for a baseline $J$ in snapshot $s$:
\begin{equation}
    \sigma^2_{sJ} = \sigma_{{\rm th},sJ}^2 + f^2|\hat{V}|^2_{sJ} + \sigma^2_{{\rm ref},sJ},
\end{equation}
where $\sigma_{{\rm th},sJ}$ is the thermal noise associated with a particular baseline and snapshot, the second term is multiplicative in the observed visibility amplitude (in order to capture residual, non-gain calibration errors, such as the aforementioned polarization leakage), and the third term is a refractive scattering noise model \texttt{J18model1} described in \citetalias{EHT2022-III}. Following \citetalias{EHT2022-II}, we fix $f = 0.02$. To mitigate diffractive scattering, the data is deblurred as in \citealt{2018ApJ...865..104J}.

We then split the data into 120\,s ``snapshots,'' flagging and removing those with fewer than four unique stations.
The visibilities in each snapshot were coherently averaged over the 120\,s window. For additional information on data processing for the snapshot fitting procedure, see \citetalias{EHT2019-III}, \citetalias{EHT2022-II}, \citetalias{EHT2022-III}, and \citetalias{EHT2022-IV}. Each snapshot contains sparse complex visibility data. Snapshots vary by ($u,v$)-coverage and the number of baselines contributing to the snapshot observation. The top row of \autoref{fig:method-schematic} shows examples of ($u,v$)-coverage for individual snapshots.

\section{Bayesian Inference in the Visibility Domain: Snapshot Fitting} \label{sec:inference}

We use the Bayesian VLBI modeling package \texttt{Comrade.jl} outlined in \citealt{Tiede2022} to form posteriors from visibility data in each snapshot. In particular, we specify a dual-cone sky emission model with the physical variables described in Section \ref{sec:model} as free parameters. We use complex visibilities instead of closure quantities to make full use of the sparse data. We therefore use an instrument model where gain amplitudes (with a Gaussian prior) and phases (with a von Mises prior) for each station present in a given snapshot are fit as free parameters.

Priors are specified in \autoref{tab:priors} for each physical parameter of interest. Since parameters of \sgra are largely unconstrained, we use wide priors in most variables. The observer's inclination relative to the spin axis is essentially unconstrained (i.e., the observer inclination is given a flat prior $\theta_o \in [1\degree, 89\degree]$, and the spin can be positive or negative). The cone opening angle is constrained to values between $40\degree$ and $90\degree$, informed by GRMHD simulations of the accretion environment of \sgra \citep{Chang2024}. Notably, we fix the mass-to-distance ratio to $5.03 \mu$as with a delta distribution. Unlike M87$^*$, the value is well constrained for \sgra \citep{2019A&A...625L..10G, 2020A&A...636L...5G, 2019Sci...365..664D}. By fixing the mass-to-distance ratio, we remove some degeneracies and aid in the constraint of other parameters.

\begin{table*}[ht!]
\centering
\caption{Physical parameter priors used for Bayesian inference on complex visibilities in each snapshot. We allow wide priors in most parameters due to the lack of constraints for \sgra, but fix the mass-to-distance ratio $\theta_{\rm g}$ based on previous studies to tighten inference on other parameters.}
\label{tab:priors}
\begin{tabular}{l|ccc}
\hline\hline
\textbf{Parameter} & \textbf{Description} & \textbf{Units} & \textbf{Prior} \\
\hline
$\theta_{\rm g}$ & Mass-to-distance ratio & $\mu$as & $\delta(5.03)$ \\
$a_{\rm *}$ & Black hole dimensionless spin & \ldots & $\mathcal{U}(-0.99, 0.99)$ \\
$\theta_{\rm o}$ & Observer inclination & rad & $\mathcal{U}(\frac{1}{180}\pi, \frac{89}{180}\pi)$ \\
$\theta_{\rm s}$ & Cone opening angle & rad & $\mathcal{U}(\frac{40}{180}\pi, \frac{\pi}{2})$ \\
$p.a.$ & Position angle of projected spin axis & rad & $\mathcal{U}(0, 2\pi)$ \\
$R_{\rm peak}$ & Characteristic radius of the emissivity profile & $\frac{GM}{c^2}$ & $\mathcal{U}(1, 10)$ \\
$p_1$ & Inner exponent of the emissivity profile & \ldots & $\mathcal{U}(0.1, 10)$ \\
$p_2$ & Outer exponent of the emissivity profile & \ldots & $\mathcal{U}(1, 10)$ \\
$\chi$ & Fluid velocity azimuthal angle in ZAMO frame & rad & $\mathcal{U}(-\pi, \pi)$ \\
$\iota$ & Magnetic field orthogonality angle in fluid frame & rad & $\mathcal{U}(0, \frac{\pi}{2})$ \\
$\beta_{\rm v}$ & Fluid speed in ZAMO frame & $c$ & $\mathcal{U}(0, 0.99)$ \\
$\sigma$ & Spectral index of emission & \ldots & $\mathcal{U}(-1, 5)$ \\
$\eta$ & Magnetic field tangential angle in fluid frame & rad & $\mathcal{U}(-\pi, \pi)$ \\
\hline
\end{tabular}
\end{table*}

\subsection{Forming Posteriors} \label{subsec:comrade}

For $N_s$ snapshots denoted by snapshot index $s$---which we approximate as independent---we fit the dual-cone model to complex visibilities, determining posterior distributions for each parameter. In particular, we fit a parameter vector $\boldsymbol{\theta}_s$. There are 13 physical parameters for the dual-cone emission model, which we denote with the parameter vector $\boldsymbol{\phi}_s$. Since we fit complex visibilities, we also include an instrument model with parameter vector $\boldsymbol{S}_s$. We define a single station gain model with a log-gain amplitude and gain phase: $g_a \equiv e^{\gamma_a + i\varphi_a}$. The log-gain amplitude is specified using a normal distribution and the gain phase with a von Mises distribution. Gain phase is calibrated relative to a reference station. 
In $\boldsymbol{S}_s$, there are two free parameters per station $I$ present in the data: $\boldsymbol{S}_s = \{\boldsymbol{\gamma}_{I,s},  \boldsymbol{\varphi}_{I,s}\}$. We jointly infer the physical parameters and instrument model $\boldsymbol{\theta}_s = \{ \boldsymbol{\phi}_s, \boldsymbol{S}_s \}$.

In a single snapshot, the joint physical parameter and instrument model posterior is given by Bayes' Theorem:
\begin{equation}
    P_s(\boldsymbol{\theta}_s | \boldsymbol{\hat{V}}_s) = \frac{\mathcal{L}_s(\boldsymbol{\hat{V}}_s|\boldsymbol{\theta}_s)\pi_s(\boldsymbol{\theta}_s)}{\mathcal{Z}_s(\boldsymbol{\hat{V}}_s)},
\end{equation}
where $\mathcal{L}_s$ denotes the likelihood, $\boldsymbol{\hat{V}}_s$ is the data available in snapshot $s$, $\pi_s$ is the prior distribution as specified in \autoref{tab:priors}, and $\mathcal{Z_s}$ is the Bayesian evidence or marginal likelihood. Following \autoref{eq:vis_obs} and \autoref{eq:vis_noise} the complex visibility likelihood is given by
\begin{equation}
    \mathcal{L}_s(\boldsymbol{\hat{V}}_s|\boldsymbol{\theta}_s) = \prod_J (2\pi \sigma^2_{s,J})^{-1} \exp[\frac{-|\boldsymbol{\hat{V}}_{s,J}-\mathcal{V}_{s,J}|^2}{2\sigma^2_{s,J}}],
\end{equation}
where $J$ is a multi-index that denotes a particular observation, time, and baseline.

A variety of degeneracies, or symmetries in the likelihood, are introduced by jointly fitting physical parameters and an instrument model, leading to non-identifiability or unconstrained degrees of freedom in the problem. We outline several prevalent examples here. 

Rescaling the flux of the image, i.e., $\tilde{I}_J \rightarrow f\tilde{I}_J$, can be absorbed by rescaling all gains: $g_a \to g_a/\sqrt{f}$. There is therefore no constraint on the total flux of sources observed with VLBI. In our analysis, we normalize the model flux to $1.0$ and renormalize the observed visibilities by the ALMA lightcurve, so that the observed visibilities on short baselines (i.e., ALMA-APEX, SMA-JCMT) are unity up to residual thermal noise effects. 

A global gain phase offset $\varphi_I \rightarrow\varphi_I + \phi_0$ has no effect on the likelihood, as the offset is canceled by its conjugate. Since likelihood is invariant under a global gain phase offset, we cannot measure the absolute phase. We infer gain phases relative to a specified reference station in each snapshot, which is selected alphabetically by station codes.

An image shift, $\tilde{I}(u,v) \to e^{2\pi i (u\Delta x + v\Delta y)}\tilde{I}(u,v)$, is a phase gradient that, similarly to gain amplitudes, can be absorbed by a global redefinition of gain phases across all stations. In VLBI, this means that the likelihood is invariant to image shifts. However, the dual-cone model sets the image origin in Bardeen coordinates to reside with the black hole, removing this degeneracy \citep{Bardeen_Cunnigham_1973}.

\subsection{Sampling Posteriors: Parallel Tempering} \label{subsec:pigeons}

We explore parameter posteriors using a sampling algorithm called \textit{parallel tempering} (PT) implemented in \texttt{Pigeons.jl} \citep{2023arXiv230809769S}. For a single snapshot, we run 20 tempering levels (separate chains) in parallel using \texttt{MPI}, with 24 threads in each tempering level task for ray-tracing the dual-cone emission model. The number of tempering levels was selected as $2\Lambda$, where $\Lambda$ is the problem's global communication barrier---around 10 for most snapshots. We perform 14 rounds of sampling, at which point the number of restarts is sufficient (greater than 100), the Bayesian evidence is stabilized, and the minimum and average swap acceptance rates $\alpha$ over the PT chains have stabilized and approached nearly the same value. We can be confident, therefore, that the sampling is robust, stable, and accurately characterizes the posterior distributions of all parameters.

During our analysis, we found that other sampling methods, such as Hamiltonian Markov Chain Monte Carlo (HMCMC) Sampling, Nested Sampling, and Dynamic Nested Sampling struggled to accurately capture the complicated, occasionally multimodal, and high-dimensional posterior in this problem. In particular, MCMC algorithms like Metropolis-Hastings and Gibbs sampling explore probability space by iteratively proposing and rejecting samples based on a target distribution. The chain may therefore struggle to traverse the low-probability regions between peaks.

The PT class of MCMC methods is capable of exploring complicated, multimodal distributions without getting stuck in a particular mode. If we are exploring a target distribution $\pi$---in this case, the posterior distribution for a Bayesian inference problem---a series of distributions of increasing complexity and sampling difficulty (an \textit{annealing path}) $\pi_1, \pi_2, ..., \pi_N$ are constructed, with $\pi_N$ equal to the target $\pi$. In an \textit{exploration phase}, samples are obtained from each distribution in parallel. In the following \textit{communication phase}, sample swaps between adjacent distributions are proposed, accepted or rejected with a Metropolis mechanism, and carried out with low communication cost \citep{10.1111/rssb.12464}. This procedure enables the discovery of new regions in the target distribution. \texttt{Pigeons.jl} implements parallel tempering with distributed computation and strong \textit{parallelism invariance}; output for a seed is identical regardless of the computational configuration, such as the number of threads, making it suitable for obtaining stable and robust scientific results \citep{2023arXiv230809769S}. For further information about the parallel tempering process used in this analysis, see \citealt{10.1111/rssb.12464}, \citealt{NEURIPS2022_03cd3cf3}, \citealt{2023arXiv230809769S}, and \citealt{pmlr-v238-biron-lattes24a}.

\subsection{Hierarchical Stacking} \label{sec:stacking}

After forming parameter posterior distributions $\boldsymbol{\theta}_s$ for all snapshots, we seek to extract the average posterior distributions to understand the persistent, stationary source structure. We build a Bayesian global/hierarchical model $\boldsymbol{\lambda} \equiv \langle\boldsymbol{\theta}_s\rangle_T$, where the $T$ indicates that we average over all timestamps. The definition of the hierarchical model implies that each of the snapshot models $\boldsymbol{\theta}_s$ can be drawn from the global distribution $\boldsymbol{\lambda}$.
We are only interested in the effects of gains on the parameter inference of the physical model and not their actual average value. 
Thus, In forming the hierarchical model, we use the distribution $P_s(\boldsymbol{\phi}_s|\boldsymbol{\hat{V}}_s) = \int P_s(\boldsymbol{\phi}_s,\boldsymbol{S}_s|\boldsymbol{\hat{V}}_s)\dd{\boldsymbol{S}_s}$, where we have marginalized over the instrument model.

We define the global posterior $P_s(\boldsymbol{\lambda} | \boldsymbol{\hat{V}}_{s})$ from the snapshot posteriors $P_s(\boldsymbol{\phi}_s | \boldsymbol{\hat{V}}_s)$ by,  

\begin{equation} \label{eq:global-post}
    P(\boldsymbol{\lambda} | \boldsymbol{\hat{V}}_{s}) = \int P(\boldsymbol{\lambda}, \boldsymbol{\phi}_s|\boldsymbol{\hat{V}}_s)\dd{\boldsymbol{\phi}_s},
\end{equation}

where the integrand is the total posterior, which is unknown. Using Bayes' rule, we rewrite the total posterior, up to a normalization constant:

\begin{equation}
    P(\boldsymbol{\lambda}, \boldsymbol{\phi}_s|\boldsymbol{\hat{V}}_s) \propto P(\boldsymbol{\hat{V}}_s | \boldsymbol{\lambda}, \boldsymbol{\phi}_s)P(\boldsymbol{\lambda}, \boldsymbol{\phi}_s).
\end{equation}

We drop $\boldsymbol{\lambda}$ in $P(\boldsymbol{\hat{V}}_s | \boldsymbol{\lambda}, \boldsymbol{\phi}_s)$ by assuming the individual snapshots do not depend on $\boldsymbol{\lambda}$ directly. Expanding the second factor:

\begin{equation}
    P(\boldsymbol{\lambda}, \boldsymbol{\phi}_s|\boldsymbol{\hat{V}}_s) \propto P(\boldsymbol{\hat{V}}_s | \boldsymbol{\phi}_s)P(\boldsymbol{\phi}_s|\boldsymbol{\lambda})P(\boldsymbol{\lambda}).
\end{equation}

We then make the reasonable approximation that each $\boldsymbol{\theta}_s$ is independent conditioned on $\boldsymbol{\lambda}$, though this is not precisely true: there is temporal correlation in reality. This allows us to decompose $P(\boldsymbol{\hat{V}}_s | \boldsymbol{\phi}_s)$ by snapshot over all $T$ timestamps and manipulate:

\begin{equation} \label{eq:total-post}
\begin{aligned}
    P(\boldsymbol{\lambda}, \boldsymbol{\phi}_s|\boldsymbol{\hat{V}}_s) &\simeq P(\boldsymbol{\lambda}) \prod_{i=1}^T P(\boldsymbol{\hat{V}}_i | \boldsymbol{\phi}_i) P(\boldsymbol{\phi}_i|\boldsymbol{\lambda}) \\
                    &= P(\boldsymbol{\lambda}) \prod_{i=1}^T \frac{P_s(\boldsymbol{\phi}_i | \boldsymbol{\hat{V}}_i)}{P(\boldsymbol{\phi}_i)}P(\boldsymbol{\phi}_i | \boldsymbol{\lambda}),
\end{aligned}
\end{equation}

where we have already specified $P(\boldsymbol{\phi}_i)$ with the priors, $\pi_s$, and the distributions $P(\boldsymbol{\phi}_i | \boldsymbol{\hat{V}}_i)$ are known through sampling on each snapshot. There is some freedom in specifying $P(\boldsymbol{\phi}_i|\boldsymbol{\lambda})$ as a modeling choice. Since we want an idea for both the average global distribution and the spread to understand variability, we parameterize persistent and variable source structure by assuming truncated unit normal/Gaussian distributions (or von Mises for cyclic parameters) in mean and standard deviation with independent parameters $p$: $P(\boldsymbol{\phi}_i|\boldsymbol{\lambda}) \sim \mathcal{TN}(\boldsymbol{\phi}_p | \boldsymbol{\mu}_p, \boldsymbol{\sigma}_p)$. Thus $\boldsymbol{\mu_p}$ will indicate the average source structure values and $\boldsymbol{\sigma_p}$ will characterize parameter variability across snapshots.

Substituting the total posterior \autoref{eq:total-post} into \autoref{eq:global-post}:

\begin{equation}
    P(\boldsymbol{\lambda} | \boldsymbol{\hat{V}}_{s}) = P(\boldsymbol{\lambda}) \prod_{i=1}^T \left( \int \frac{P_s(\boldsymbol{\phi}_i | \boldsymbol{\hat{V}}_i)}{P(\boldsymbol{\phi}_i)}P(\boldsymbol{\phi}_i | \boldsymbol{\lambda}) \dd{\boldsymbol{\phi}_i} \right).
\end{equation}

The integral is implicitly approximated with our sampling of the posterior $P(\boldsymbol{\phi}_i | \boldsymbol{\hat{V}}_i)$. With $N$ samples:

\begin{equation}
    P(\boldsymbol{\lambda} | \boldsymbol{\hat{V}}_{s}) = P(\boldsymbol{\lambda}) \prod_{i=1}^T \frac{1}{N} \left( \sum_{i=1}^N \frac{P_s(\boldsymbol{\phi}_i | \boldsymbol{\lambda})}{P(\boldsymbol{\phi}_i)} \dd{\boldsymbol{\phi}_i} \right)
\end{equation}

Finally, we perform parallel tempered MCMC sampling of the global average mean posterior and variability posterior. Each sample drawn from the global posterior is a Gaussian (or von Mises) distribution.

It is important to note that our stacking approximation overspecifies the priors for the hierarchical model. 
The overspecification is evident from considering Bayes' rule,
\begin{equation}
    P(\boldsymbol{\theta}_i|\boldsymbol{\lambda})P(\boldsymbol{\lambda}) = P(\boldsymbol{\lambda}|\boldsymbol{\theta}_i)P(\boldsymbol{\theta}_i),
\end{equation}
where we have specified $P(\boldsymbol{\theta}_i|\boldsymbol{\lambda})$ with a normal distribution $\mathcal{N}$ and $P(\boldsymbol{\mu})$ with a uniform distribution $\mathcal{U}(\mu_l, \mu_h)$. 
From these considerations, we can analytically solve for $P(\boldsymbol{\theta}_i)$ directly:
\begin{equation}
    P(\boldsymbol{\theta}_i) = \int P(\boldsymbol{\theta}_i|\boldsymbol{\lambda}) P(\boldsymbol{\lambda}) d\boldsymbol{\lambda}
\end{equation}
However, we have selected a uniform prior for each snapshot $P(\boldsymbol{\theta}_i) = \mathcal{U}(l, h)$, overspecifying the model. This procedure is nonetheless suitable for our purposes.

\section{Results} \label{sec:results}

\subsection{Posteriors from a Single Snapshot} \label{subsec:snapshot}

To demonstrate snapshot fitting, we include the results of fitting one snapshot of \sgra data, indexed as scan 157 (April 7, 2017 at 13:49 UTC). \autoref{fig:scan157-mean-std-img} shows the mean and standard deviation images produced by a series of images drawn from the sampled parameter posteriors, alongside a subset of those varied sample images. The mean image displays expected features of optically-thin plasma accretion flow in a Kerr metric, such as an inner shadow, diffuse direct ($n=0$) emission, and a bright photon ring, with different brightness asymmetries in the $n=0$ and $n=1$ sub images. The data in a single snapshot is limited, resulting in unconstrained parameters and multimodality; there is therefore wide variation among images which properly fit the data, as shown in the standard deviation image. Because it aggregates a variety of image morphologies, the mean image is not indicative of the real physical sky brightness distribution.

The full corner plot showing the joint marginal probability densities for all pairs of physical parameters is shown in \autoref{fig:scan157-full-pairplot} (aside from the mass-to-distance ratio, which is fixed). While several parameters are relatively well constrained, others, especially the spin, are unconstrained. In this snapshot, a low position angle and close to face-on observer inclination are preferred. The position angle is multimodal, with a wrap-around at $2\pi$ and another mode at $\pi$. This snapshot favors a cone opening angle close to $40\degree$ rather than an opening angle closer to equatorial. The characteristic emission radius is well constrained due to the fixed mass-to-distance ratio eliminating degeneracies. 
Most parameters have support across the entire range. The results from this snapshot draw on sparse data and are subject to the temporal variability of \sgra: results for a single snapshot likely do not capture persistent source structure.

\subsection{Hierarchical Model Posteriors} \label{subsec:hypermodel}
\begin{figure*}[!ht]
    \centering
    \includegraphics[width=\textwidth]{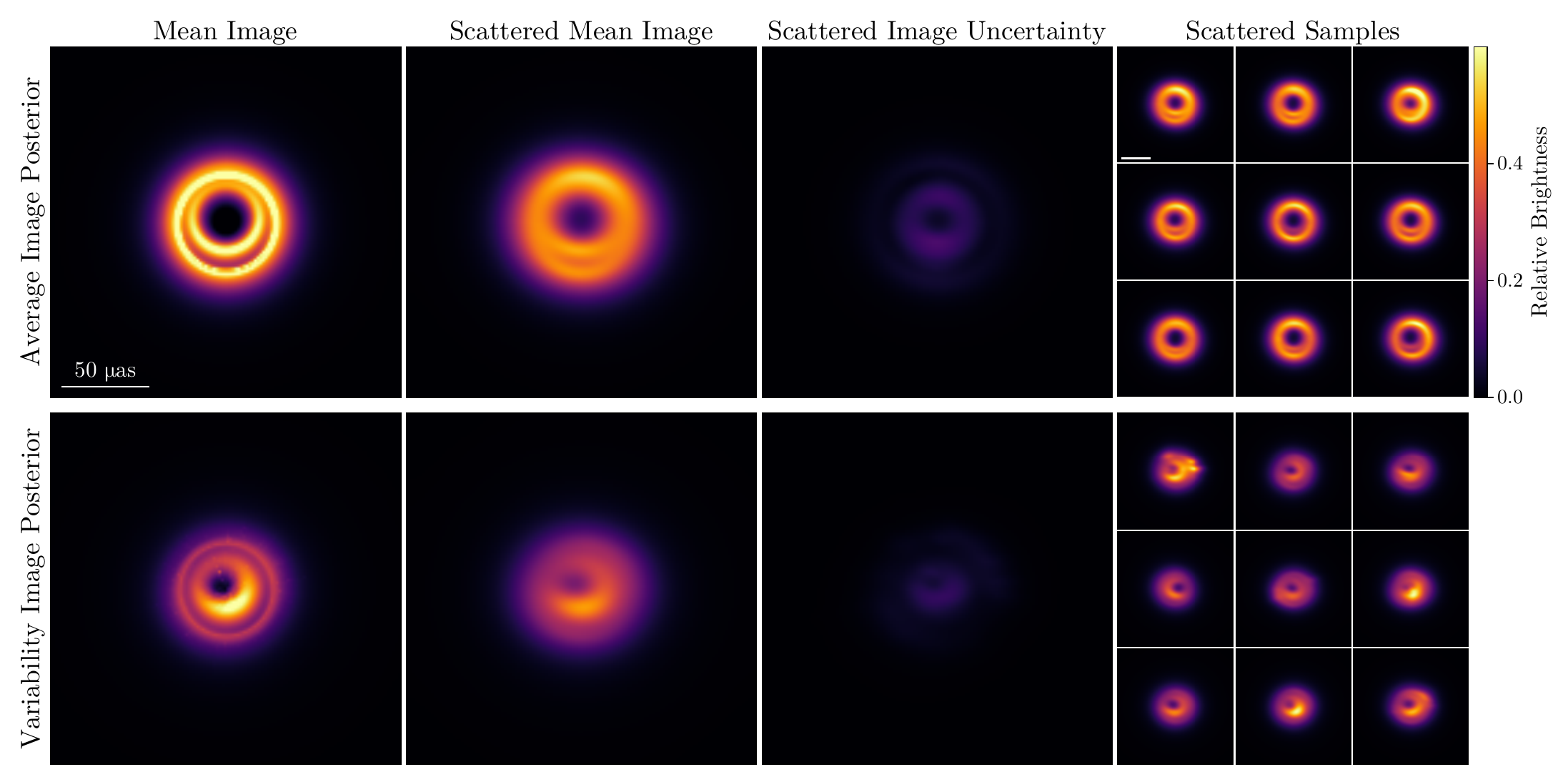}
    \caption{Images produced from \sgra hierarchical averaging results for both the average source structure distribution $\boldsymbol{\mu}$ (top row) and the variability distribution $\boldsymbol{\sigma}$ (bottom row). The first column shows the mean of image samples, the second column the scattered mean image, the third column the scattered standard deviation or uncertainty about the mean, and the fourth column samples from the marginal hierarchical posterior. Scattered images are produced by blurring the nominal image with the diffractive scattering kernel from \citet{2018ApJ...865..104J}. The mean image from the average source structure posterior is highly symmetric, with a visible inner shadow and a bright photon ring. Individual image draws in the average posterior display less variable morphologies than image samples for a single snapshot posterior; the standard deviation image suggests this variation is confined primarily to near-horizon emission interior to the photon ring. The mean variability map displays the images features most subject to variability across snapshots.}\label{fig:global-img}

\end{figure*}

Using the procedure described in Section \ref{sec:stacking}, we form a Bayesian hierarchical model, or \textit{hypermodel}, from the posterior chains produced by sampling each of the $161$ snapshots analyzed. Gaussian distributions of mean $\boldsymbol{\mu}_p$ and width, or standard deviation, $\boldsymbol{\sigma}_p$, are sampled from the resulting global posterior. We refer to the mean $\boldsymbol{\mu}_p$ as the average source structure distribution and the standard deviation $\boldsymbol{\sigma}_p$ as the variability distribution. We use flat priors for the average source structure and variability distributions---von Mises for angular, wrapping variables and uniform otherwise. Parameter values are constrained to the same prior range listed in \autoref{tab:priors}. We perform hierarchical stacking with no restrictions on the possible variability of each parameter across snapshots. 

\begin{figure*}[!htb]
    \centering
    \subfigure(a){\includegraphics[width=0.45\textwidth]{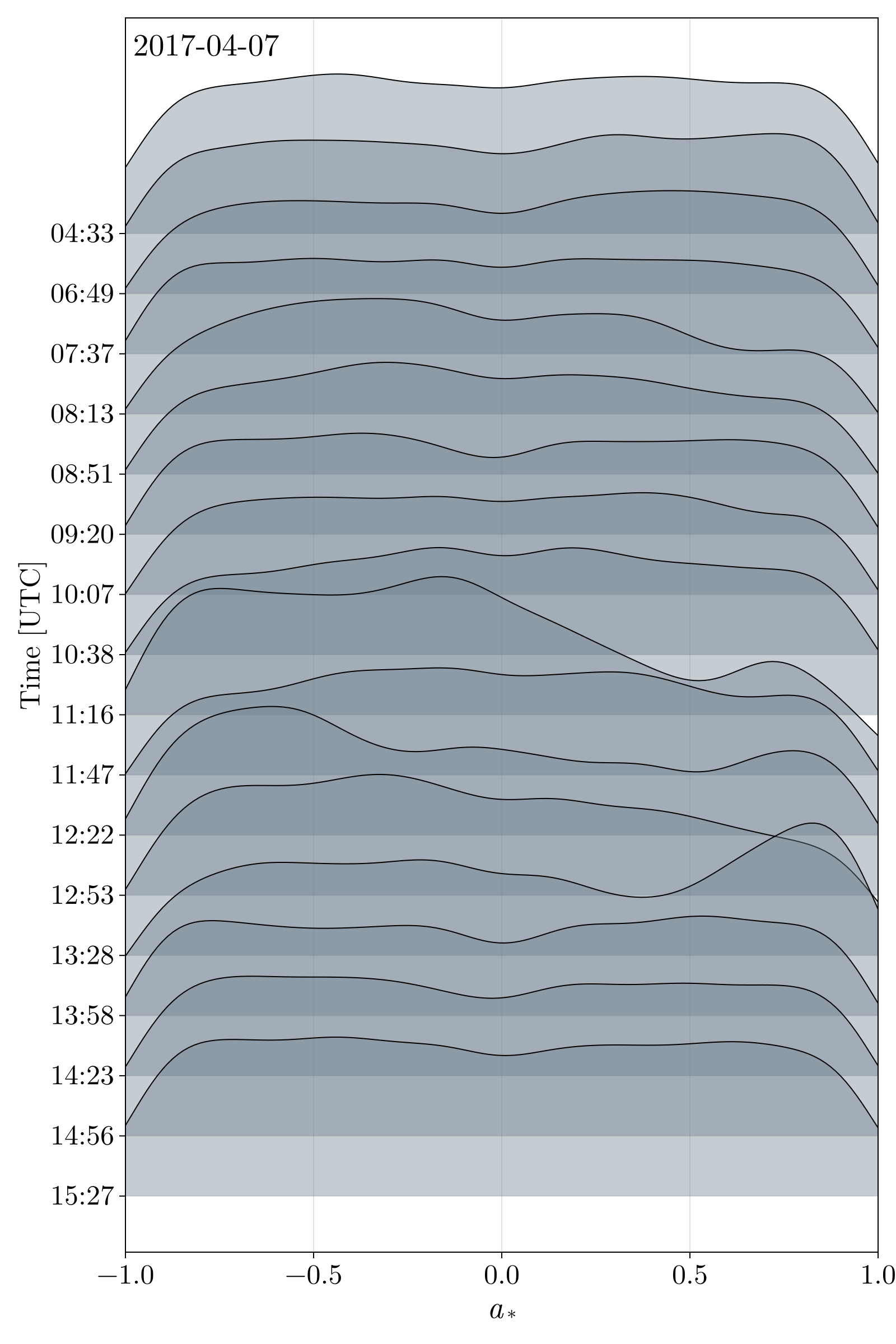}} 
    \subfigure(b){\includegraphics[width=0.45\textwidth]{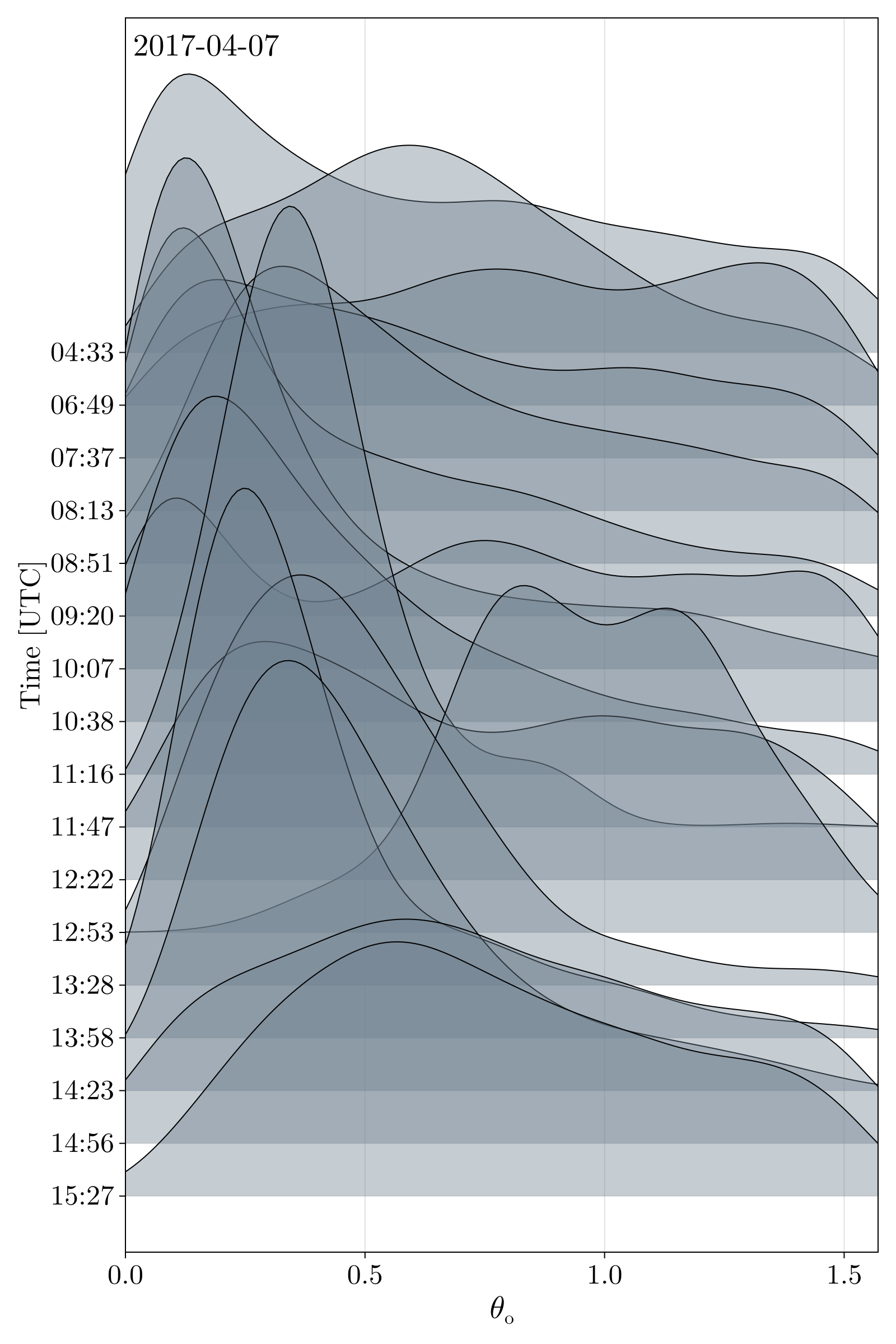}} 
    \caption{Snapshot inference applied to the \sgra 2017 data set. The parameter posteriors from every tenth snapshot are stacked vertically. Results for (a) the black hole spin parameter $a_{\rm *}$ and (b) the observer inclination $\theta_{\rm o}$ are shown. Spin posteriors are wide, with occasional preference towards positive or negative spin in some snapshots. Inclination posteriors, however, show a strong and stable tendency for lower observer inclinations across snapshots.}
    \label{fig:ridgeplots}
\end{figure*}

\begin{figure*}[ht!]
\centering
\includegraphics[width=0.95\linewidth]{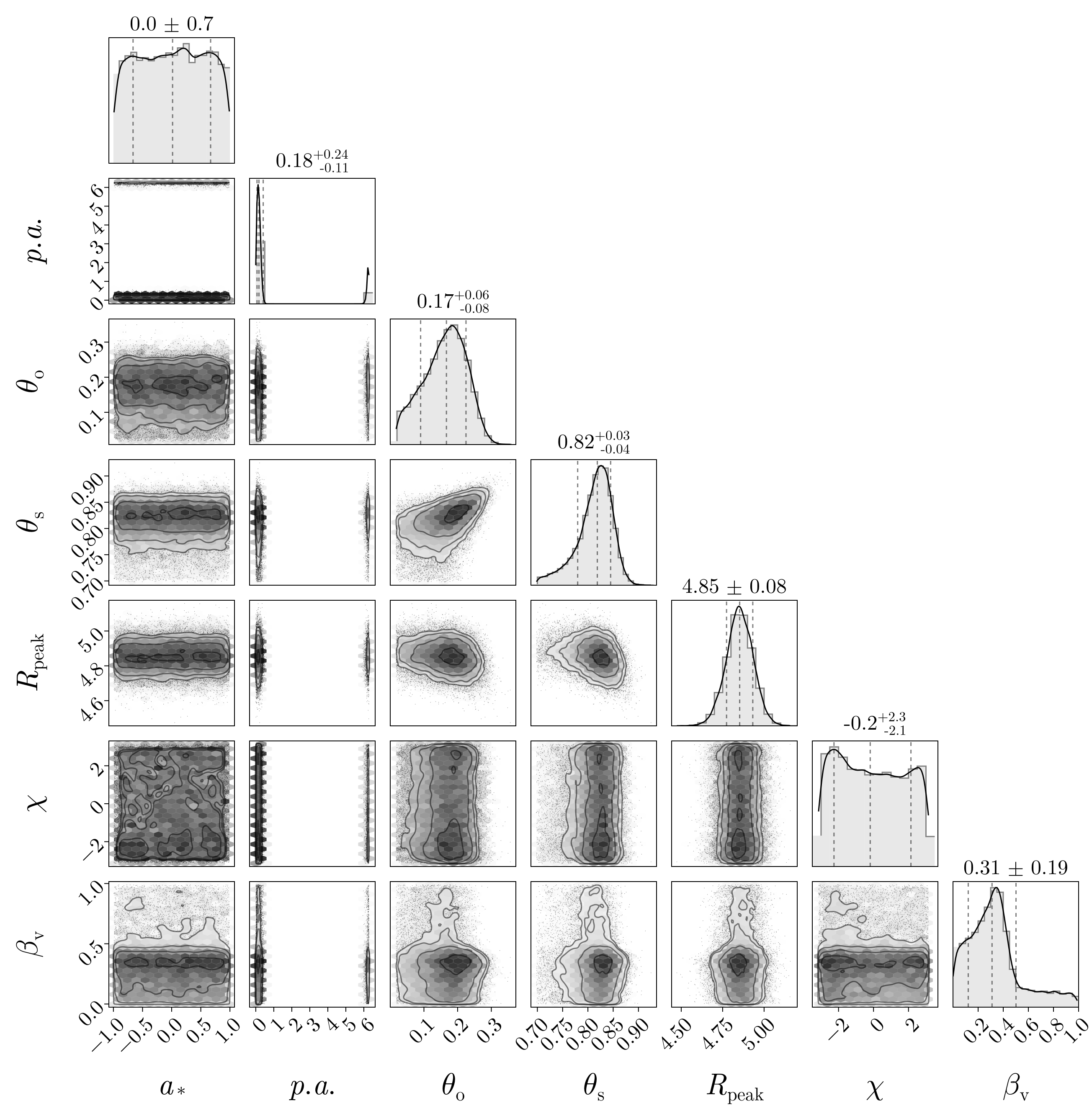}
\caption{Joint marginal probability densities for a subset of the averaged model parameters. The black hole spin and azimuthal angle of the fluid flow $\chi$ are unconstrained. The spin position angle, black hole spin inclination, cone opening angle, and characteristic emission radius are well constrained by our analysis. Up to model specification, the hierarchical model suggests a nearly face-on observer inclination, a nearly vertical spin axis, strong near-horizon emission, dominant non-equatorial emission, and slow-moving plasma relative to the innermost stable circular orbit (ISCO). 
\label{fig:global-pairplot-trunc}}
\end{figure*}

Several parameters, like black hole spin $a_{\rm *}$, $\eta$, and $\iota$ are poorly constrained by this analysis. Other parameters of interest, however, are more tightly constrained, up to model specification. The mean, mean with scattering applied, and standard deviation/uncertainty images for the hierarchical model average source structure distribution $\boldsymbol{\mu}_p$ are shown in \autoref{fig:global-img} alongside a series of scattered sample images drawn from $\boldsymbol{\mu}_p$. To produce scattered images, nominal images are blurred with the diffractive scattering kernel from \citet{2018ApJ...865..104J}. \autoref{fig:global-img} also displays an estimate for a mean variability map with uncertainty and image samples. To make the variability map, we use around 1600 samples from the hierarchical model. For each sample, we create a Gaussian distribution (or von Mises for angular parameters) from the mean and variability values in each parameter. We draw 200 values from these distributions, create 200 images, and take the standard deviation of the images. The mean of the standard deviation images across all $\sim 1600$ samples produces the variability map, and the standard deviation of these images is the uncertainty of the variability map. Relative brightness values are shown, because we unit normalize image flux for analysis. Statistical quantities of both the average source structure $\boldsymbol{\mu}_p$ and variability $\boldsymbol{\sigma}_p$ posterior distributions are listed in \autoref{tab:stacking-no-restrict}. \autoref{fig:global-pairplot-trunc} shows the corner plot from a subset of parameters of interest; \autoref{fig:global-full-pairplot} shows the full corner plot for all physical parameters. Subsets of black hole spin and observer inclination posteriors varying over time are shown in \autoref{fig:ridgeplots}.

\begin{table}[ht!]
\centering
\caption{Parameter values inferred by the Bayesian hierarchical model. $\langle\boldsymbol{\mu}_p\rangle$, $\sigma_{\boldsymbol{\mu}_p}$, and  $\widetilde{\boldsymbol{\mu}_p}$ are the mean value, standard deviation, and median, respectively, of a given parameter $p$ in the mean source structure distribution $\boldsymbol{\mu}$. Distances from the median to the lower quantile at $16\%$ $\widetilde{\boldsymbol{\mu}_p}-$ and the upper quantile at $84\%$ $\widetilde{\boldsymbol{\mu}_p}+$ are shown alongside the median. The circular mean and standard deviation are presented for angular variables $p.a.$, $\chi$, and $\eta$. The same values are displayed for parameters in the variability distribution $\boldsymbol{\sigma}$.}
\label{tab:stacking-no-restrict}
\begin{tabular}{l|cccc}
\hline\hline
\textbf{Par.} & \textbf{$\langle\boldsymbol{\mu}_p\rangle$} $\pm$ \textbf{$\sigma_{\boldsymbol{\mu}_p}$} & $\widetilde{\boldsymbol{\mu}_p}$ & \textbf{$\langle\boldsymbol{\sigma}_p\rangle$} $\pm$ \textbf{$\sigma_{\boldsymbol{\sigma}_p}$} & $\widetilde{\boldsymbol{\sigma}_p}$ \\
\hline
$a_{\rm *}$ & $0.01\pm 0.57$ & $0.02^{+0.65}_{-0.68}$
 & $10.7\pm5.2$ & $10.8^{+6.1}_{-6.2}$
 \\
$\theta_{\rm o}$ & $0.16\pm0.06$ & $0.17^{+0.05}_{-0.08}$
 & $0.20 \pm 0.03$ & $0.20^{+0.04}_{-0.03}$
 \\
$\theta_{\rm s}$ & $0.81\pm0.04$ & $0.82^{+0.03}_{-0.04}$
 & $0.10 \pm 0.02$ & $0.10^{+0.03}_{-0.02}$
 \\
$p.a.$ & $0.13\pm0.12$ & $0.18^{+0.23}_{-0.11}$ & $0.76 \pm 0.08$ & $0.75^{+0.09}_{-0.07}$
 \\
$R_{\rm peak}$ & $4.85\pm0.08$ & $4.85^{+0.08}_{-0.08}$ & $0.50 \pm 0.07$ & $0.51^{+0.06}_{-0.07}$
 \\
$p_1$ & $4.93\pm2.85$ & $4.90^{+3.39}_{-3.28}$
 & $53.2 \pm 26.7$ & $53.2^{+31.3}_{-31.1}$
 \\
$p_2$ & $4.75\pm0.13$ & $4.75^{+0.12}_{-0.13}$
 & $0.68 \pm 0.09$ & $0.67^{+0.1}_{-0.08}$
 \\
$\chi$ & $-2.37 \pm 2.38$ & $-0.19^{+2.33}_{-2.07}$
 & $31.5 \pm 18.1$ & $31.5^{+21.5}_{-21.4}$
 \\
$\iota$ & $0.82 \pm 0.45$ & $0.83^{+0.51}_{-0.55}$ & $8.32 \pm 4.27$ & $8.35^{+4.97}_{-5.03}$
 \\
$\beta_{\rm v}$ & $0.33 \pm 0.22$ & $0.31^{+0.19}_{-0.19}$
 & $2.10 \pm 2.76$ & $0.54^{+4.99}_{-0.24}$
 \\
$\sigma$ & $-0.69 \pm 0.21$ & $-0.72^{+0.26}_{-0.19}$ & $0.94 \pm 0.15$ & $0.93^{+0.15}_{-0.14}$
 \\
$\eta$ & $0.86 \pm 2.35$ & $0.18^{+1.91}_{-2.21}$ & $31.4 \pm 18.2$ & $31.5^{+21.5}_{-21.7}$
 \\
\hline
\end{tabular}
\end{table}

\subsection{Inference on Synthetic GRMHD Data}\label{subsec:grmhd-inference}

\begin{figure*}[!ht]
    \centering
    \includegraphics[width=\linewidth]{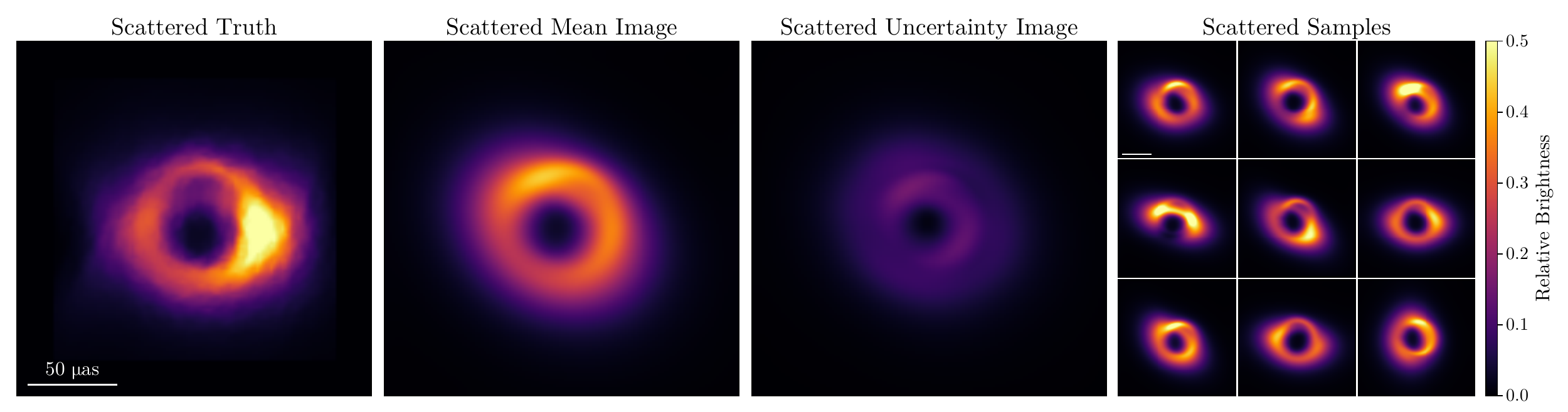}
    \caption{Hierarchical averaging image results for the GRMHD synthetic data test. Left to right: the true average image used to construct the data, including the impact of both diffractive blurring and refractive scintillation;  mean image from the average source structure posterior, blurred with the diffractive scattering kernel for \sgra; the scattered image uncertainty/standard deviation; samples from the average source structure posterior. While the mean posterior image angular structure differs from the ground truth, the posterior image structure is highly uncertain, and some samples are similar to the true image.}
    \label{fig:grmhd_image_comp}
\end{figure*}

To test the reliability of our results, we performed the snapshot modeling pipeline on synthetic data generated from a GRMHD simulation with known ground truths in several parameters of interest. The Magnetically Arrested Disk (MAD) GRMHD simulation has spin $a_*=-0.5$, inclination $\theta_o=50\degree$, position angle $p.a.=178\degree$, and mass-to-distance ratio $\theta_g=5.563\mu$as (we use this as the prior on $\theta_{\rm g}$ and keep the other priors consistent with the \sgra analysis). Thermal electron population temperatures are modeled with $R_{high}=40$, a parameter which sets the ratio of ion to electron temperatures \citep[see \citetalias{EHT2019-V, EHT2022-V} and][]{2016A&A...586A..38M}. Complex visibilities were simulated to emulate the \sgra April 7, 2017 observing window. The $(u,v)$-coverage in the simulated snapshot visibilities is identical to coverage in the \sgra data used in this paper. \autoref{fig:grmhd_image_comp} shows the scattered and refracted true time-averaged GRMHD image along with the average structure distribution scattered mean image, scattered uncertainty map, and a subset of scattered samples showing the rich diversity of morphologies arising from our inference. Posteriors for the spin, inclination, and position angle from the average source structure distribution in the synthetic data set are shown in \autoref{fig:grmhdtest-pairplot-truth}. \autoref{fig:grmhd-full-pairplot} in Appendix \ref{sec:grmhd-appendix} shows all parameter posteriors.

\begin{figure}[ht!]
\centering
\includegraphics[width=0.95\linewidth]{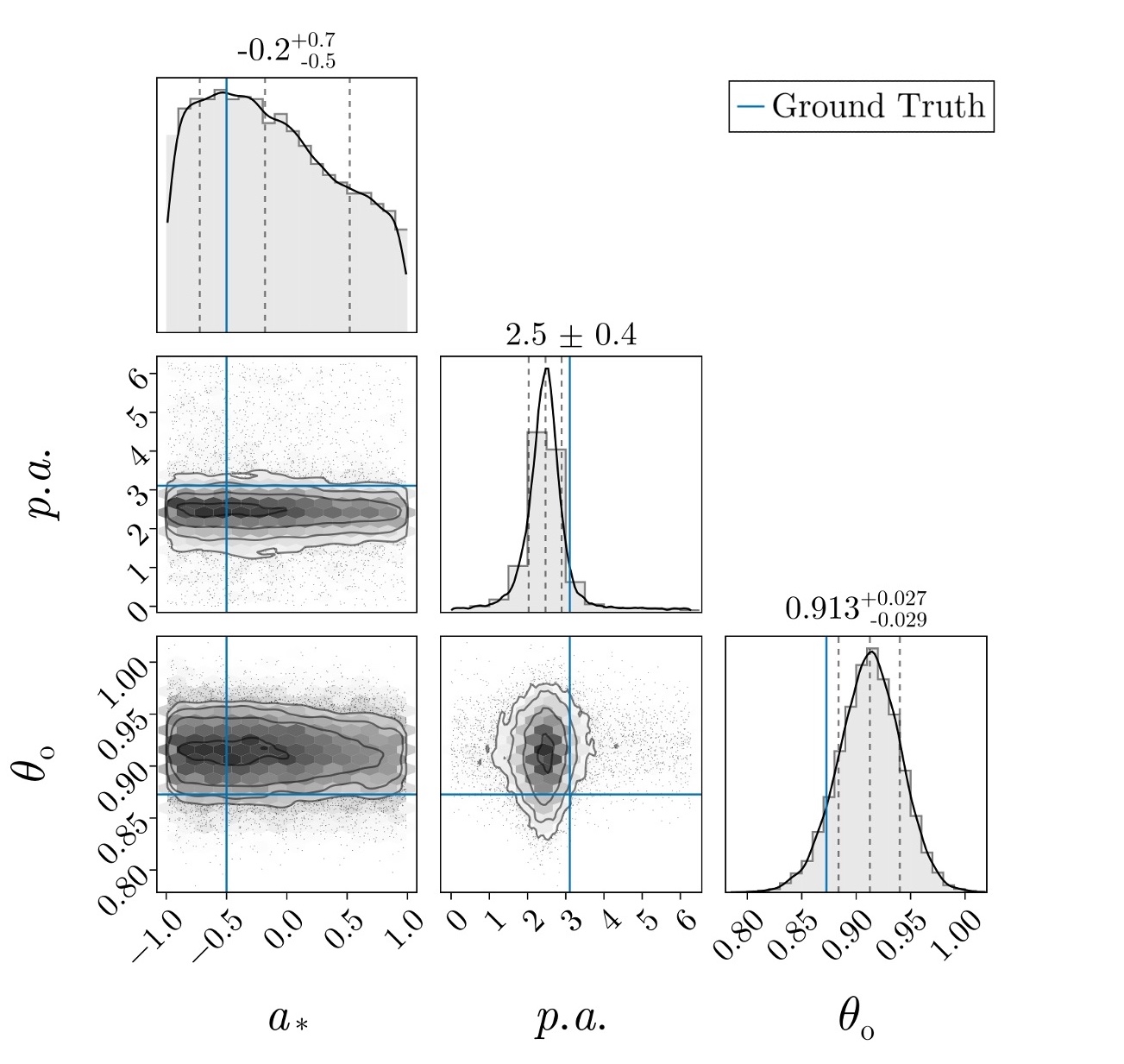}
\caption{Partial pair plot from the average source structure distribution of the global hierarchical model formed from snapshot modeling synthetic data. Ground truth values are known for the spin, observer inclination, and position angle; truth lines are shown in blue. The spin position angle and observer inclination are shown in radians. Spin is unconstrained, corroborating our unconstrained spin result for \sgra visibility data, though the posterior peaks around the true spin value. The inference nearly recovers both the true spin position angle and observer inclination, revealing no significant biases in inference of black hole spin, spin position angle, or observer inclination.
\label{fig:grmhdtest-pairplot-truth}}
\end{figure}

As in our \sgra visibility results, the inference pipeline is unable to confidently recover black hole spin in the synthetic data without additional model assumptions and constraints. However, the black hole spin posterior peaks near the true spin value and a negative spin is correctly inferred: $a_{\rm *}=-0.12\pm 0.54$. The true position angle is nearly recovered, though it is variable across snapshots and not very tightly constrained: $p.a. = 144\degree \pm43 \degree$. The inferred observer inclination is also near ground truth and more tightly constrained: $\theta_{\rm o}=52.3\degree \pm1.7\degree$. We therefore find no significant biases on spin, inclination, or position angle in our inference pipeline. A slow plasma speed $\beta_{\rm v}$ and near-equatorial emission are also inferred, generally consistent with the MAD simulation.

\section{Discussion} \label{sec:discussion}

There are important limitations and assumptions to consider when analyzing the results of the Bayesian hierarchical model. In the dual-cone emission model, the plasma and magnetic field are axisymmetric about the equatorial plane. Even if the persistent source structure and plasma geometry of \sgra is axisymmetric, there may be non-axisymmetric features in individual snapshots. When fitting non-axisymmetric data to an axisymmetric model, the model infers parameter distributions which conform to emission geometry asymmetries and irregularities. This introduces degeneracies in the posterior distributions and complicates inference. Since the data is sparse, however, this may not be of tremendous concern for individual snapshot inference.

Nonetheless, consequences of fitting nonstationary, non-axisymmetric, highly time variable accretion flows with an axisymmetric model are more apparent in the global posterior formed via hierarchical stacking without variability restrictions. Black hole spin is unconstrained across snapshots, with nearly uniform distributions and support across the entire prior, suggesting spin measurements with the 2017 EHT array are infeasible without further assumptions and data products. The mean spin value from the source structure distribution is near zero with large uncertainty. The mean variability in black hole spin across snapshots, however, is significant. Physically, we expect the spin to remain stable and persistent across the snapshots used in this analysis. The spin variation is a result of the lack of spin constraints and is likely indicative of model misspecification, which was expected. In this case, time-dependent features of the accretion flow may be absorbed into variation in the spin and possibly other parameters. This misspecification---though reduced in effect by the sparsity of data---is important to consider when forming conclusions about the physical environment of \sgra. As discussed in Section \ref{subsec:comrade}, the dual-cone model assumes a specific origin for the black hole, assuaging image translation, centroid, and gain phase degeneracies.

Our inference produces mean images in each snapshot and in a hierarchical model with expected features of accretion flows around spinning black holes, like the shadow and photon ring. In the images produced from a given snapshot, such as scan 157 (April 7, 2017 at 13:49 UTC), we see significant variation in image morphologies due to the lack of parameter constraints with sparse visibility coverage (see \autoref{fig:method-schematic} for examples of ($u,v$)-coverage); this variation is apparent in the standard deviation/uncertainty image and sample image draws in \autoref{fig:scan157-mean-std-img}. There is less, but still some, morphology variation in images produced using the Bayesian hierarchical model. The mean image of the average source structure distribution is nearly symmetric, in agreement with previous EHT studies \citepalias{EHT2022-I} and with studies of the morphology of \sgra at 3.5\,mm wavelength \citep{2019ApJ...871...30I}. The standard deviation image of the average source structure distribution, shown in \autoref{fig:global-img}, indicates that much of this variation is prominent within the direct and indirect emission interior to the photon ring. The variability map in \autoref{fig:global-img} displays strong intraday variability in near-horizon emission.

While images give us an idea about the possible morphologies which fit the snapshot data, the primary interests of this analysis are physical parameter posteriors. In \autoref{fig:scan157-full-pairplot}, we see that a single snapshot with good coverage (around 9 baselines) is partially capable of constraining a few parameters and incapable of constraining others. Here, we consider each parameter---how posteriors appear in a single snapshot, across all snapshots, and in the Bayesian hierarchical model---and the possible physical implications of the results.

The black hole spin of \sgra is a parameter of particular interest to the EHT. Constraining spin is an ongoing and prevalent problem in the field of black hole imaging; a black hole's spin has significant implications for the spacetime structure surrounding the black hole, emission properties, jet formation processes, and many other areas of active research \citep[e.g.][]{1977MNRAS.179..433B, EHT2019-V, 2022MNRAS.511.3795N}. Previous EHT studies of \sgra tend to slightly favor positive black hole spin and low inclination, but variable intrinsic source structure, the potential for missing physical aspects in the emission models, and discrepancies between the observed visibilities and GRMHD simulations prevented conclusive constraints \citepalias{EHT2022-V, EHT2024-VIII}. 

In this analysis, the spin is highly unconstrained, with slight variations across snapshots; a few snapshots tend towards positive spin (spin axis pointing toward the observer). \autoref{fig:ridgeplots} shows several black hole spin posteriors over different time stamps/snapshots. In the average source structure distribution of the Bayesian hierarchical model, the spin is near-zero with significant uncertainty (nearly uniform distributions with support across the entire prior), consistent with the lack of spin constraint across snapshots (see \autoref{tab:stacking-no-restrict} and \autoref{fig:global-pairplot-trunc}). As discussed above, there is also significant variability in the average spin across snapshots. This result corroborates the difficulty of obtaining accurate black hole spin measurements. Observations with increased $(u,v)$ sampling may improve our ability to make accurate, constraining spin measurements, but further work is required to develop physics-based models that reduce degeneracies which are absorbed into the spin. Black hole spin has a more pronounced effect on the photon ring (the $n=1$ and greater sub images) than on the direct emission, so a more extensive ground-based radio telescope network may still struggle to constrain spin via the methodology used in this paper, motivating the addition of a space-based telescope to the network. Inference with polarization data may also help constrain spin, as frame-dragging around the black hole affects observed polarization patterns \citep{2020ApJ...894..156P, Palumbo_2022, Chael_2023}.

The observer inclination angle is tightly constrained by this analysis. In most snapshots and in the average source structure distribution, fitting the data with the dual-cone emission model favors a face-on observer inclination, nearly looking down the angular momentum axis. Inference consistently prefers low inclination angles (see \autoref{fig:ridgeplots}). Temporal variability is present, but not highly significant. A nearly face-on observer inclination for \sgra agrees with previous studies adopting different analysis methods \citep[\citetalias{EHT2022-V, EHT2024-VIII}; see also][]{2018A&A...615L..15G}. In particular, ``best-bet'' models based on \sgra constraints were found to have low inclination angles \citepalias{EHT2022-V}.

The cone opening angle is unconstrained in many snapshots but ultimately tends towards lower values $\sim 45\degree$, indicating significant non-equatorial emission arising from, e.g., a thick accretion disk, spherically infalling material, or emission associated with a jet.  The cone opening angle is a model-specific parameter and not a direct physical parameter, so implications of this result are strongly dependent on model specification. Inferred cone opening angles in several snapshots approach the GRMHD-informed lower bound; inference on several snapshots with a wider cone opening angle prior did not significantly change the results, however. There are feasible scenarios in which \sgra is more optically thick than expected; an optically thin model such as the dual-cone emission model used in this analysis would therefore favor face-on inclination with a low cone opening angle. Indeed, the \sgra Spectral Energy Decomposition (SED) suggests the source may be on the boundary between optically thick and optically thin \citep{Broderick_2016}. Additionally, \citealt{Broderick_2016} produced similar images to the mean image of the average source structure distribution in \autoref{fig:global-img} using optically thick accretion models with a line-of-sight inclination around 60$\degree$ with respect to the spin axis.

Notably, the dual-cone emission model is more reasonable for face-on than edge-on systems. Edge-on viewer inclinations result in image streaks, an artifact of the thin emission surfaces in the model. In the future, the model can be adapted for optically thick, edge-on scenarios with a consideration of optical depth effects and a scale height for emission surfaces. Such three-dimensional emission modeling would require radiative transfer effects, including absorption processes like synchrotron self-absorption, or an optical depth prescription for each crossing of emission surfaces.

The spin position angle is tightly constrained, with two modes arising from wrapping of the angular variable. The hierarchical model constrains the position angle tightly, though the degree to which the variable is constrained in each snapshot varies. Other modes around $\pi$ arise in some snapshots. There is historically significant uncertainty in characterization of the dynamics of \sgra, and position angle may evolve over the April 7, 2017 observing period, affecting the inferred persistent position angle \citepalias{EHT2022-III}. 
Both $\chi$ and $p.a.$ contribute to brightness asymmetries in the image. This degeneracy and the lack of a spin constraint may contribute to the poor constraint of $\chi$.

The characteristic emission radius $R_{\rm peak}$ is also well constrained in most snapshots and in the average source structure distribution. Our inference suggests near-horizon emission dominates the \sgra source. 

The inner index for the emissivity profile power law, $p_1$, is more poorly constrained than the outer index, $p_2$ and exhibits significant variability, suggesting that direct, near-horizon emission varies in intensity over time. A well-constrained outer index---which largely impacts emission near the photon ring---is not surprising, as previous EHT results constrain ring width decently well \citepalias{EHT2022-IV}. The mean $p_1$ and $p_2$ from the source structure distribution are comparable in value, suggesting emission within and beyond the characteristic radius grows and falls off, respectively, at a similar rate. A negative spectral index of emission is favored, that implies---if a power law electron distribution is assumed---a rising spectrum not typical of synchrotron emission from optically thin plasma. Additionally, \citealt{2022ApJ...930L..19W} found a slightly positive, near-zero typical spectral index for \sgra using lightcurves gathered during the EHT 2017 observing campaign. However, our analysis involves only one frequency and this result may be indicative of slight misspecification of the accretion geometry or electron distribution. There is also strong support for near zero spectral index values. The discrepancy indicates that inference with an optically thick plasma model may be necessary to properly capture the source data \citep[e.g.][]{Broderick_2016, 2018ApJ...863..148P}.

The fluid velocity azimuthal angle $\chi$ is well constrained in some snapshots, unconstrained in others, and often multimodal: since spin is largely unconstrained, a particular sign of $\chi$ could indicate prograde or retrograde flows in any given sample. The average source structure distribution of the hierarchical model also displays this multimodality in $\chi$; the median value less than around $-\pi /2$ corresponding to the stronger mode indicates radially inward flows. Previous EHT studies have found that \sgra emission is consistent with models using Radiatively Inefficient Accretion Flows (RIAFs)/Advection Dominated Accretion Flows (ADAFs), in which radially inward flows within the innermost stable circular orbit (ISCO) are supported. \citep[\citetalias{EHT2022-V, EHT2024-VIII}]{2018ApJ...863..148P}. Much of the $\chi$ posterior density lies in the region of radially inward flows (confined to the conic emission regions of the model), though $\chi$ is not constrained well enough for conclusive estimates. In a cut of samples with only positive spin, the $\chi$ posterior slightly favors radially inward and retrograde flows, but prograde flows are still strongly supported. Radially inward accretion flows typically produce bright direct emission, but to a lesser extent than radially outward flows. We find high time variability in $\chi$, suggesting that the particular direction of fluid flow varies widely over observations. This may indicate turbulence in the accreting magnetized plasma, consistent with expected flows in LLAGN \citepalias{EHT2019-V}. The uncertainty on the temporal variability is large, however, so the nature of the variability is inconclusive.

The magnetic field orthogonality angle $\iota$ and tangential magnetic field angle are poorly constrained. Both magnetic field configuration parameters---especially the azimuthal angle in the fluid frame $\eta$--- exhibit high variability. Snapshot modeling with a consideration of light polarization may provide greater insight into the magnetic field configuration.

Our inference more confidently suggests plasma accretion flows with relatively low speed---less than half the speed of light---both in individual snapshots and in the global average structure distribution. This may not directly correspond to the speed of the plasma in \sgra because the model fluid is confined to flow within conic regions. At lower speeds, direct emission remains more prominent in the resulting image. Additionally, Doppler boosting is weakened in slowly rotating models \citepalias{EHT2022-V}. There is still posterior support for faster speeds up to $0.99c$.

Inferencing on synthetic visibilities from a GRMHD simulation showed no significant biases in inferred black hole spin, position angle, or observer inclination. This result confers confidence on the independent \sgra observer inclination measurement in this paper, up to model limitations. The position angle obtained from the synthetic data notably less tightly constrained than the \sgra position angle. Previous semi-analytic studies have shown that the spin position angle can be significantly biased by the state of accretion flow \citetalias{EHT2022-III}. Though synthetic data inference reveals no significant bias for the spin position angle, the distinctly different inferred posteriors suggest further analysis is needed for confidence in the \sgra spin position angle measurement. Myriad image morphologies arise from the inferred average source structure parameter posteriors (see \autoref{fig:grmhd_image_comp}). The mean image differs from the ground truth, time-averaged GRMHD image, as it is a representation of the statistical spread of possible morphologies and not a precise image reconstruction.

Previous EHT studies found that jet-dominated models of \sgra match observational constraints only at small viewing angles, while disk-dominated models match observational constraints more broadly \citep{2019ApJ...871...30I}. The low observer inclination inferred in this analysis supports the possibility of jet-dominated emission. This analysis may benefit from using an equatorial plane emission surface and a diffuse jet-like emission component in addition to the conical surfaces for future exploration. Additionally, we limit our analysis to Stokes I intensity images; considering light polarization may be of interest in the future. Using more datasets, especially when more stations are involved in observations, will help to further constrain parameters. In this analysis we are using real observational data. It would be informative to apply this process to simulated observations with the expected next generation Event Horizon Telescope (ngEHT) station configuration. Greater $(u,v)$-coverage would ease posterior sampling and may enable tighter parameter constraints. Inference on simulated data with a variety of refractive scattering models applied is of interest for improving the scattering models used in processing pipelines. Finally, performing inference with models that include neglected effects in our analysis such as finite optical depth and multi-component emission (arising from potential large-scale diffuse structures, like jets) would reduce (potentially severe) biases from model misspecification.

\section{Conclusions} \label{sec:conclusions}

Parameter inference for \sgra is complicated by significant temporal variability, with source structure varying on the order of minutes to hours. Reconstructing images and physical characteristics of the accretion environment around the SMBH requires novel approaches which do not rely on traditional Earth rotation aperture synthesis for improving VLBI coverage. In this analysis, we perform Bayesian inference on 120\,s ``snapshots'' of sparse complex visibility data from the EHT's April 7, 2017 observation campaign of \sgra. We directly fit ray-traced, semi-analytic, dual-cone accretion flow models in the Kerr spacetime to visibilities, jointly inferring both physical model parameters and instrument models/complex gains for each telescope station. Posteriors are explored via parallel tempered MCMC sampling. We then combine individual snapshot posteriors into a single global model through Bayesian hierarchical stacking.

An important limitation in this analysis is potential model misspecification introduced by fitting nonstationary, non-axisymmetric accretion flows with an axisymmetric model. There are also important physical processes and additional degrees of freedom in \sgra which our models do not account for. Up to model specification, however, we infer values and uncertainties for physical parameters and characterize their intraday variability.
Our models display strong statistical support for a nearly face-on observer inclination ($\theta_o = 9.2\degree \pm 3.6 \degree (\pm_{\rm v} 11.6\degree \pm_{\rm v} 2.0\degree)$), an emission peak near the horizon ($R_{peak} = 4.9 \pm 0.1(\pm_{\rm v} 0.5\pm_{\rm v} 0.1) \,GM/c^2$), near-vertical projected spin position angle ($p.a. = 7.3\degree \pm 7.08 \degree (\pm_{\rm v} 43.5\degree \pm_{\rm v} 4.6\degree)$), and dominant emission $43.4\degree \pm 2.0\degree (\pm_{\rm v} 5.9\degree \pm_{\rm v} 1.4\degree)$ above the equatorial plane, where the first uncertainties are the modeling error and values in parentheses indicate temporal variability and its uncertainty. Our inference also favors slow-moving plasma flows relative to the ISCO ($\beta_v = 0.33\pm 0.22 (\pm_{\rm v} 2.1\pm_{\rm v} 2.8)$), though posterior support remains for higher speeds. Even with a fixed mass-to-distance ratio and many physical assumptions, the spin and magnetic field configuration parameters are largely unconstrained. Inference on GRMHD synthetic visibilities revealed no significant biases in black hole spin, position angle, or observer inclination measurements. We therefore provide an independent measure of the observer inclination relative to the black hole spin axis of \sgra, up to model limitations.

This work provides plausible constraints to inform our understanding of the characteristics and environment of \sgra. During future observational campaigns and theoretical analysis of \sgra through full, ray-traced GRMHD or semi-analytic accretion models, we may employ these inferred parameter values for improved physical understanding. If we can constrain $\chi$, for instance, we would likely improve our understanding of $a_{\rm *}$, or vice versa. Our snapshot modeling pipeline can be easily applied to visibilities in all past and future EHT \sgra observation campaigns, enabling tighter parameter constraints and the characterization of interday and interyear variability. Parameters with more significant temporal variability likely require multi-epoch observations and inference on data over longer timescales for sufficient constraints.

In the future, we intend to mitigate the effects of model misspecification. We will explore extended models---such as a model with both dual-cone and equatorial disk emission, or other multi-component emission models to account for potential large-scale, diffuse structures---and incorporate effects like non-axisymmetric fluid flows which may more successfully capture \sgra's turbulent, time variable flows. Since the cone opening angle posteriors often approach the lowermost bound on the prior, we will likely expand that prior for future analysis. Our current accretion model assumes an optically thin plasma. The possibility of a typical $n = 0$ optical depth greater than one for \sgra, however, necessitates considering an optically thick semi-analytic model. Alternatively, adding an optical depth prescription to each crossing of emission surfaces in the model would likely improve applicability to \sgra and smooth some sharp-edged features resulting from viewing the cones at certain observer inclinations.

Further model expansion includes using emission surfaces with scale heights for more accuracy in reproducing edge-on images and considering radiative transfer effects; such additions would need to be computationally efficient enough for inference. The snapshot modeling approach currently assumes snapshots are independent and uncorrelated, which is not physically true. To consider the correlation between snapshots, inference on all snapshots would need to be run in parallel on GPU's. Lastly, inference could be extended to fitting polarized light, where polarization calibration effects and improvements in the dual-cone emission model polarization signatures would need significant attention. Performing this analysis on more \sgra snapshot data---especially in observation campaigns with additional telescopes---would also allow us to more effectively constrain parameters and long term variability.

Two major observational limitations that affect the snapshot fitting are 1) the sparsity of data in each snapshot, and 2) the limited angular resolution, which cannot distinguish direct emission from the photon ring. The addition of new telescopes to the EHT will steeply increase the number of instantaneous baselines \citep{EHT_Mid-Range,ngEHT_Key_Science,Doeleman_2023}, while planned extensions to ground-space baselines through the Black Hole Explorer \citep[BHEX;][]{BHEX_Concept,BHEX_Instrument,BHEX_Photon_Ring} would isolate emission from the photon ring. Future studies should also explore extensions including modeling the refractive scattering directly and exploring the implications of different scattering models on the model inference.

\begin{acknowledgments}

Braden J. Marazzo-Nowicki is funded through generous support from Mr.\ Michael Tuteur and Amy Tuteur, MD.
We acknowledge financial support from the National Science Foundation (AST-2307887). This publication is funded in part by the Gordon and Betty Moore Foundation, Grant GBMF-12987. This work was supported by the Black Hole Initiative, which is funded by grants from the John Templeton Foundation (Grant \#62286) and the Gordon and Betty Moore Foundation
(Grant GBMF-8273)---although the opinions expressed in this work are those of the author(s) and do not necessarily reflect the views of these Foundations. This work used the RCAC Anvil Cluster at Purdue University through allocation PHY250202 from the Advanced Cyberinfrastructure Coordination Ecosystem: Services \& Support (ACCESS) program \citealt{10.1145/3569951.3597559}, which is supported by U.S. National Science Foundation grants \#2138259, \#2138286, \#2138307, \#2137603, and \#2138296.
    
\end{acknowledgments}




\appendix

\section{Semi-Analytic Accretion Flow Model}\label{sec:model-appendix}

\begin{figure*}[!htb]
    \centering
    \subfigure{\includegraphics[width=0.24\textwidth]{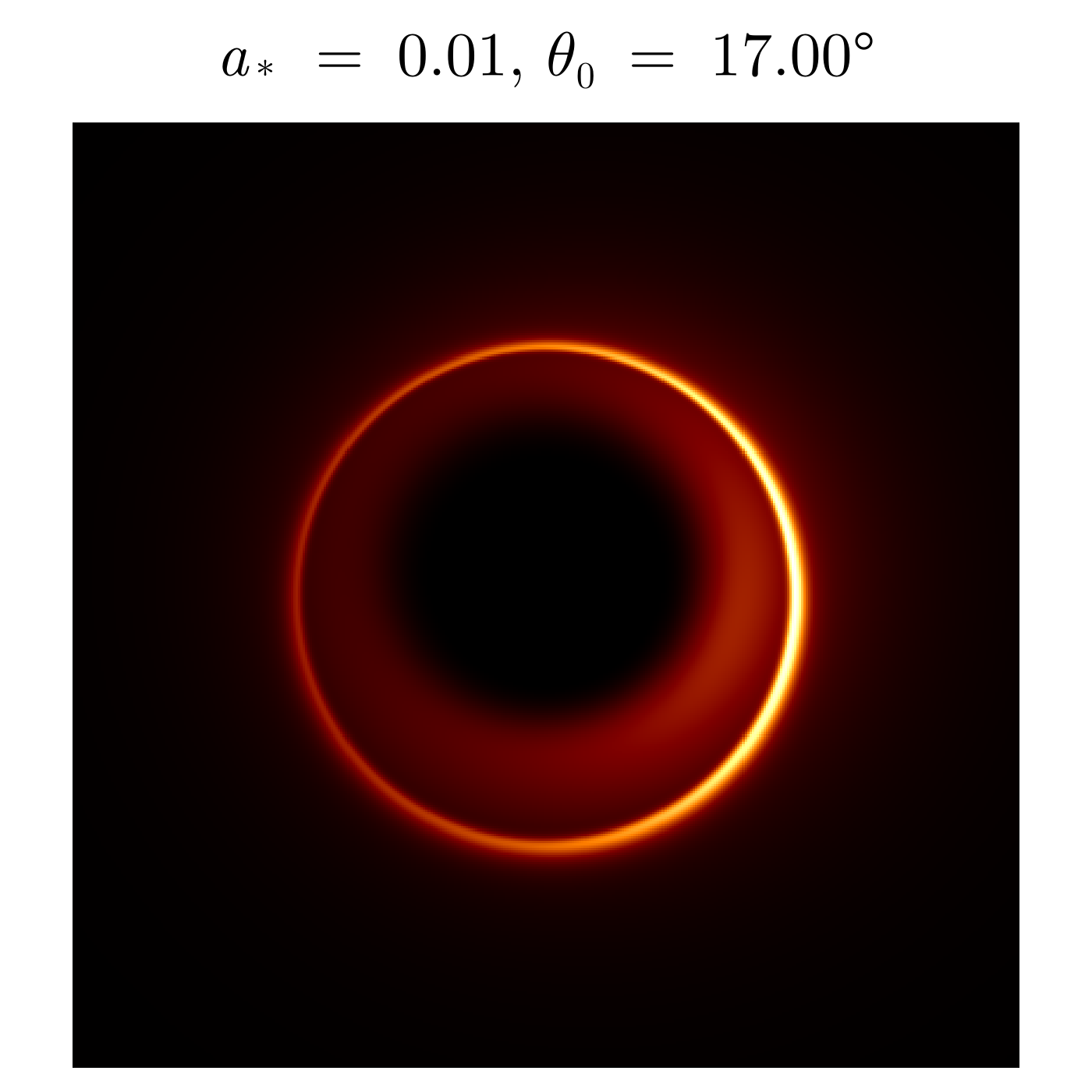}} 
    \subfigure{\includegraphics[width=0.24\textwidth]{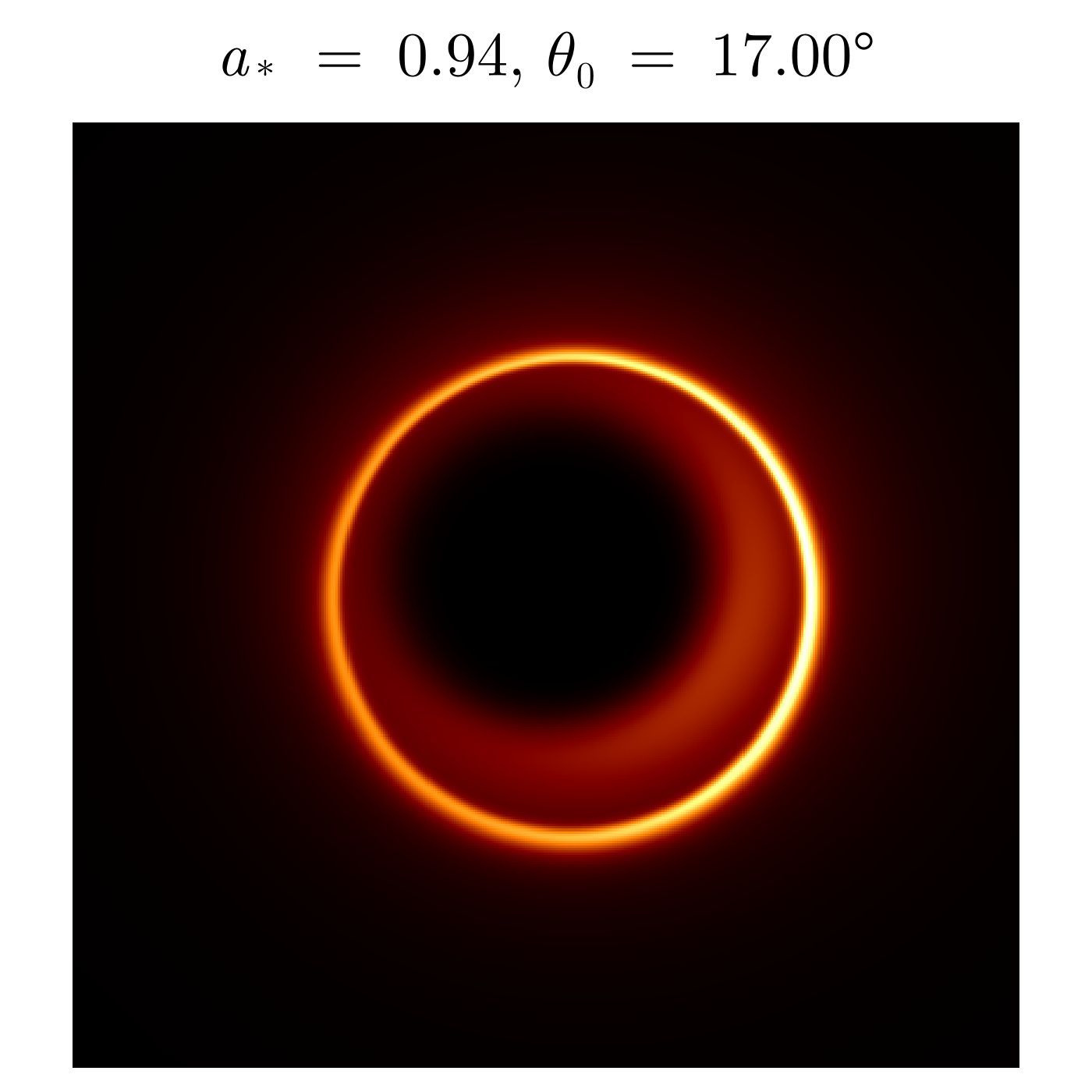}} 
    \subfigure{\includegraphics[width=0.24\textwidth]{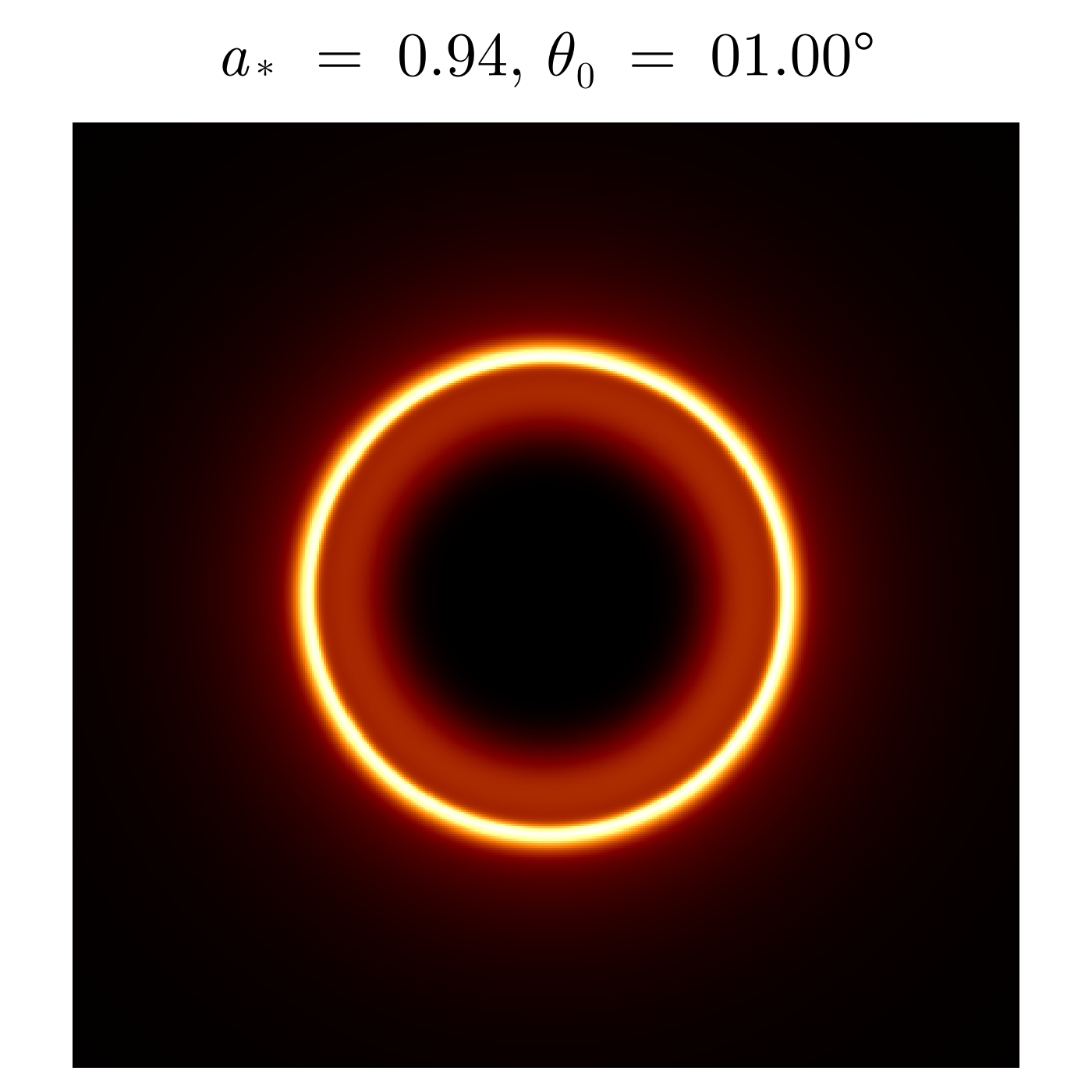}} 
    \subfigure{\includegraphics[width=0.24\textwidth]{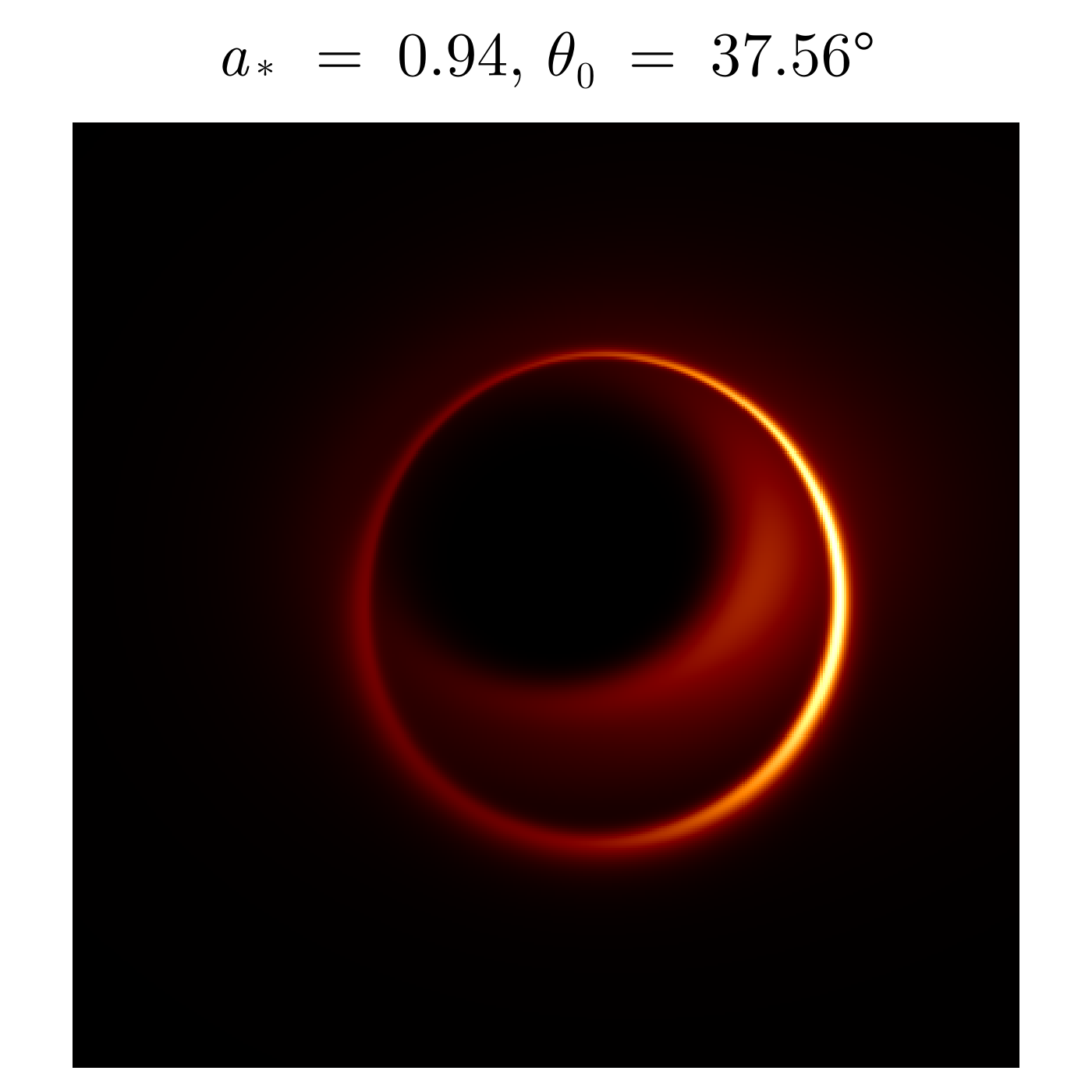}} 
    \subfigure{\includegraphics[width=0.24\textwidth]{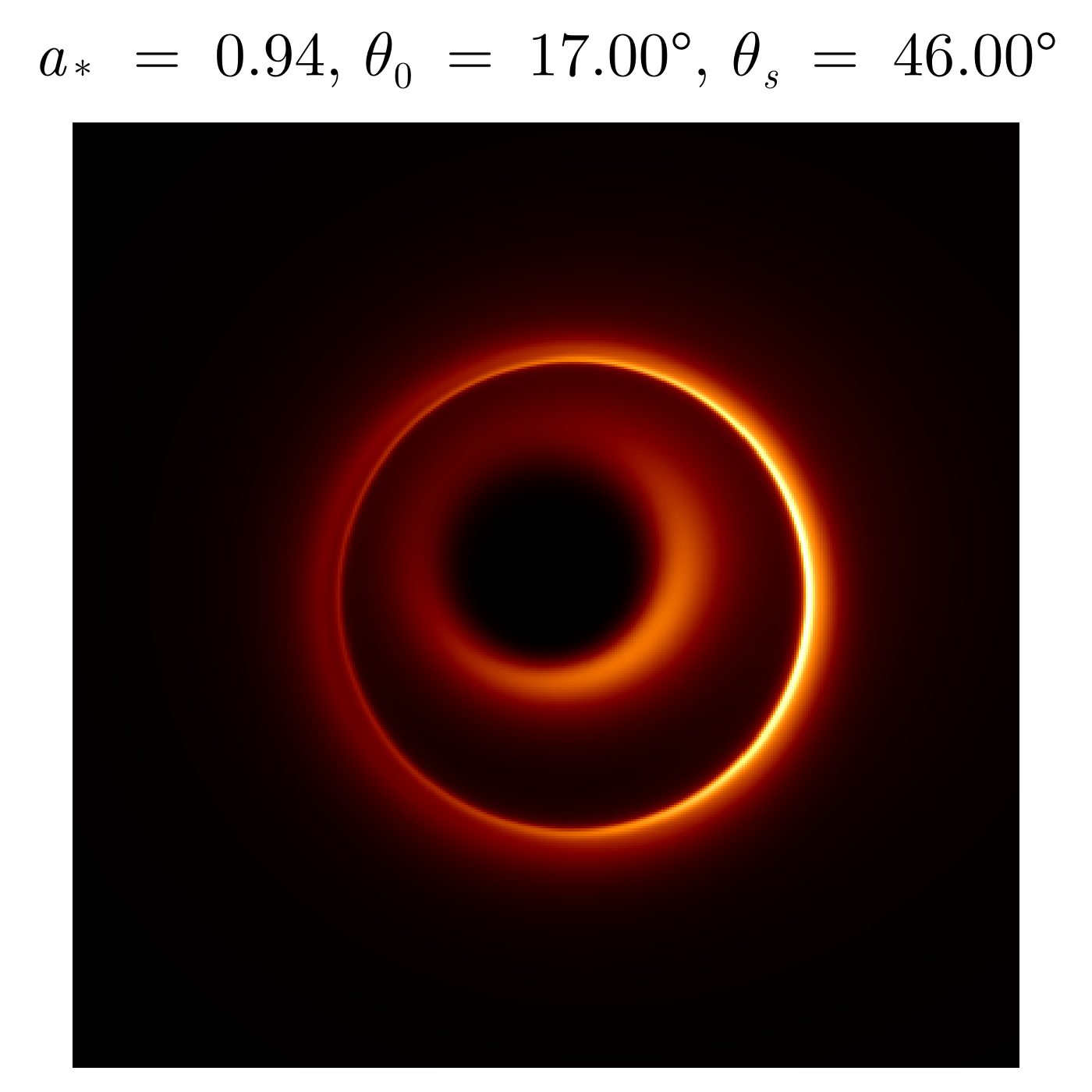}}
    \subfigure{\includegraphics[width=0.24\textwidth]{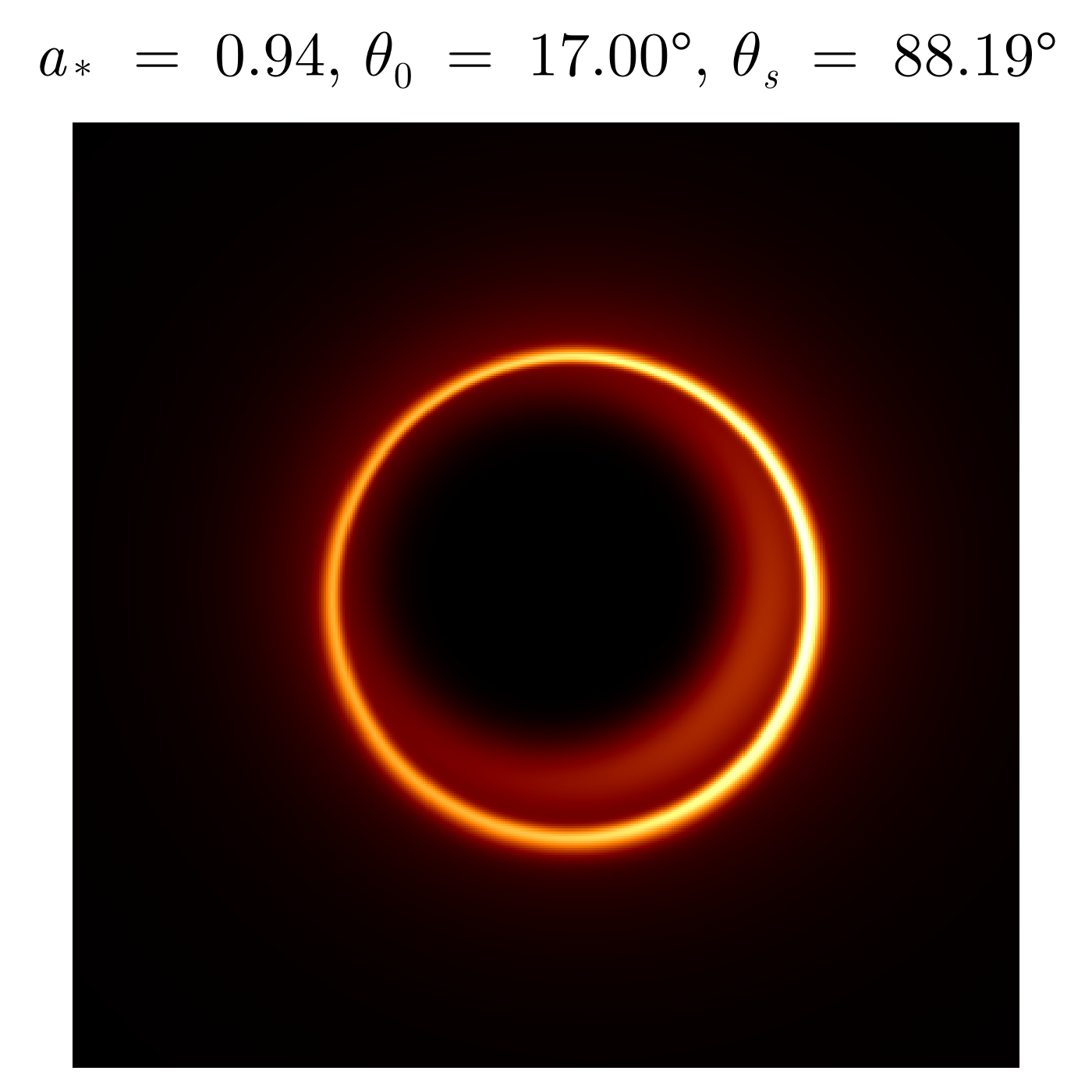}}
    \subfigure{\includegraphics[width=0.24\textwidth]{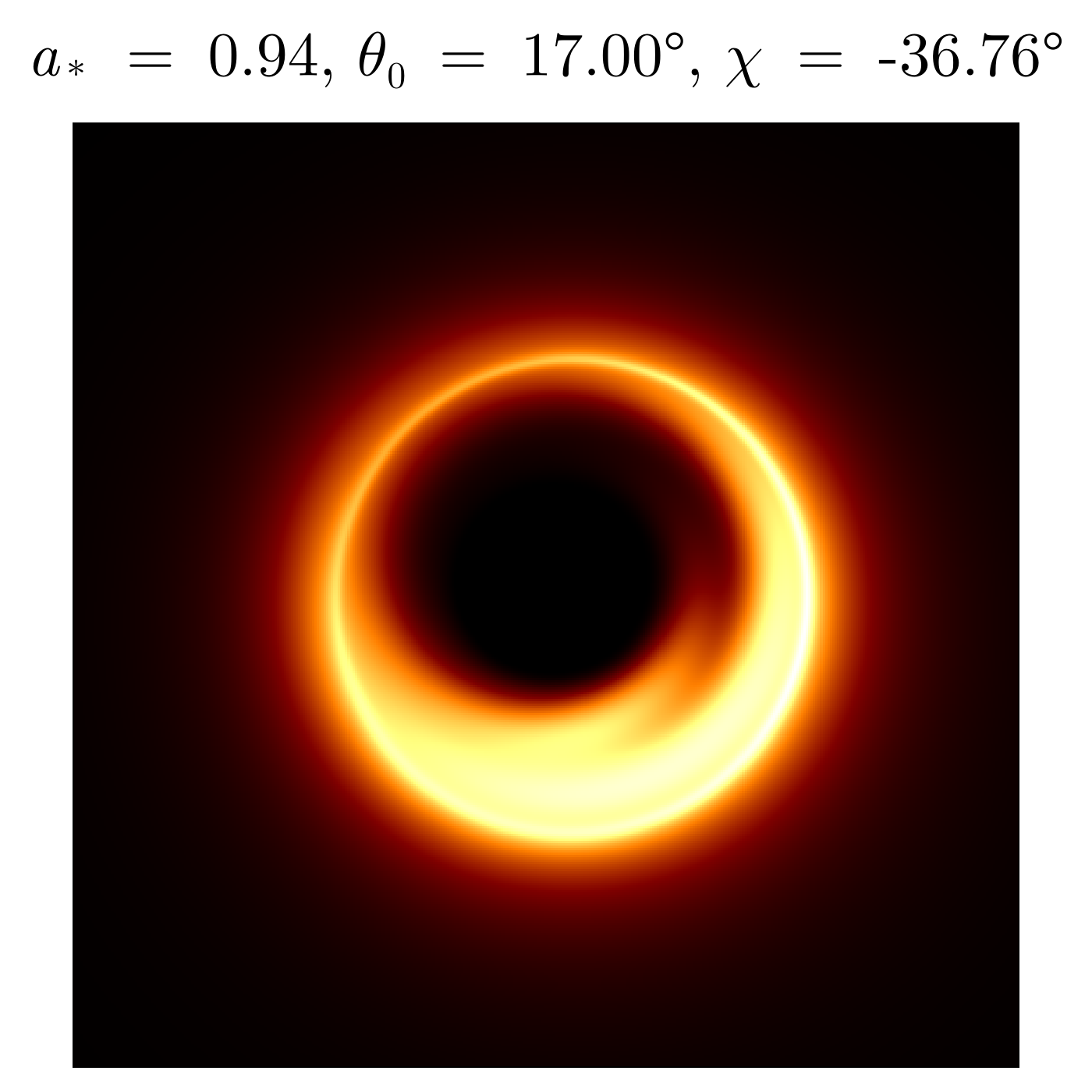}}
    \subfigure{\includegraphics[width=0.24\textwidth]{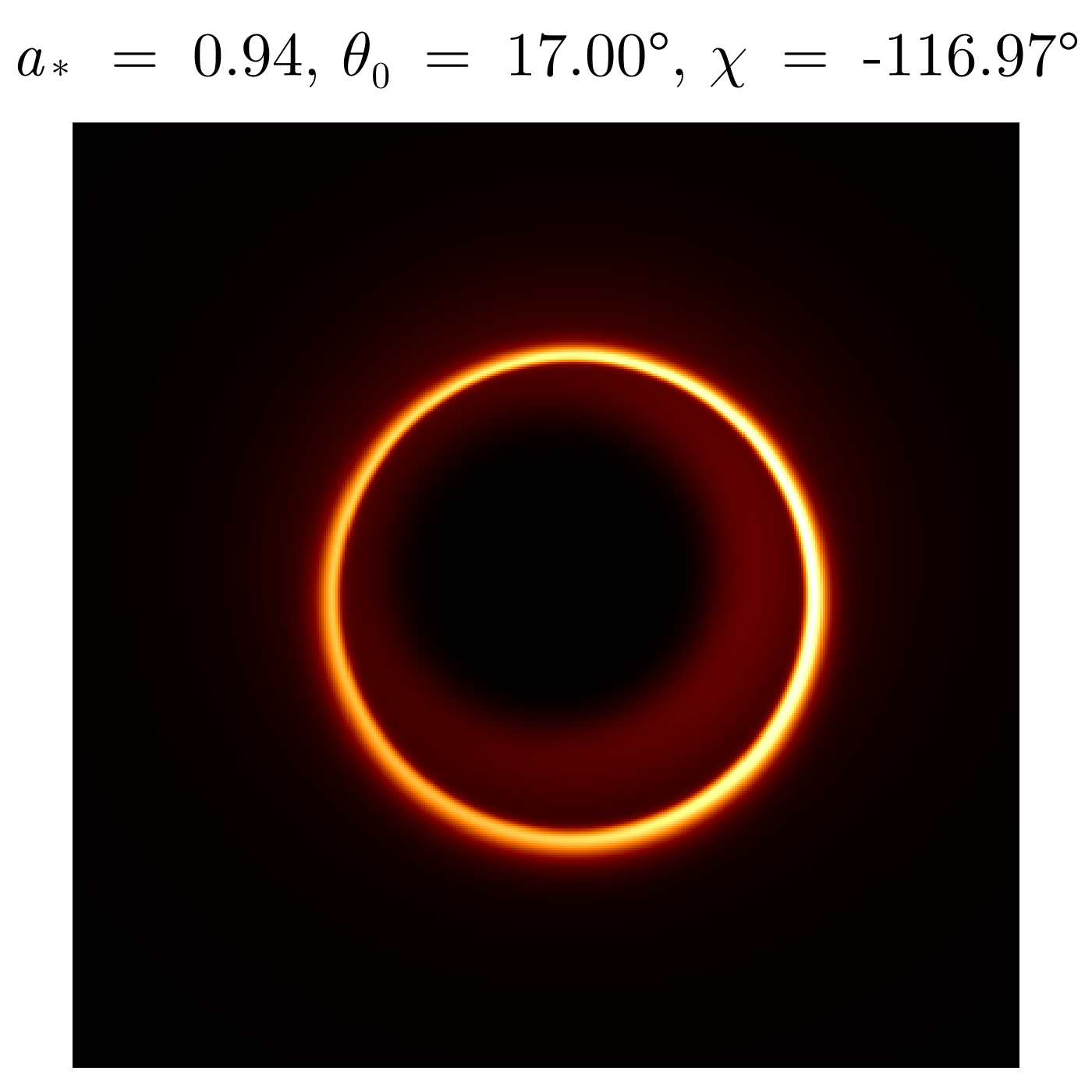}}
    \subfigure{\includegraphics[width=0.24\textwidth]{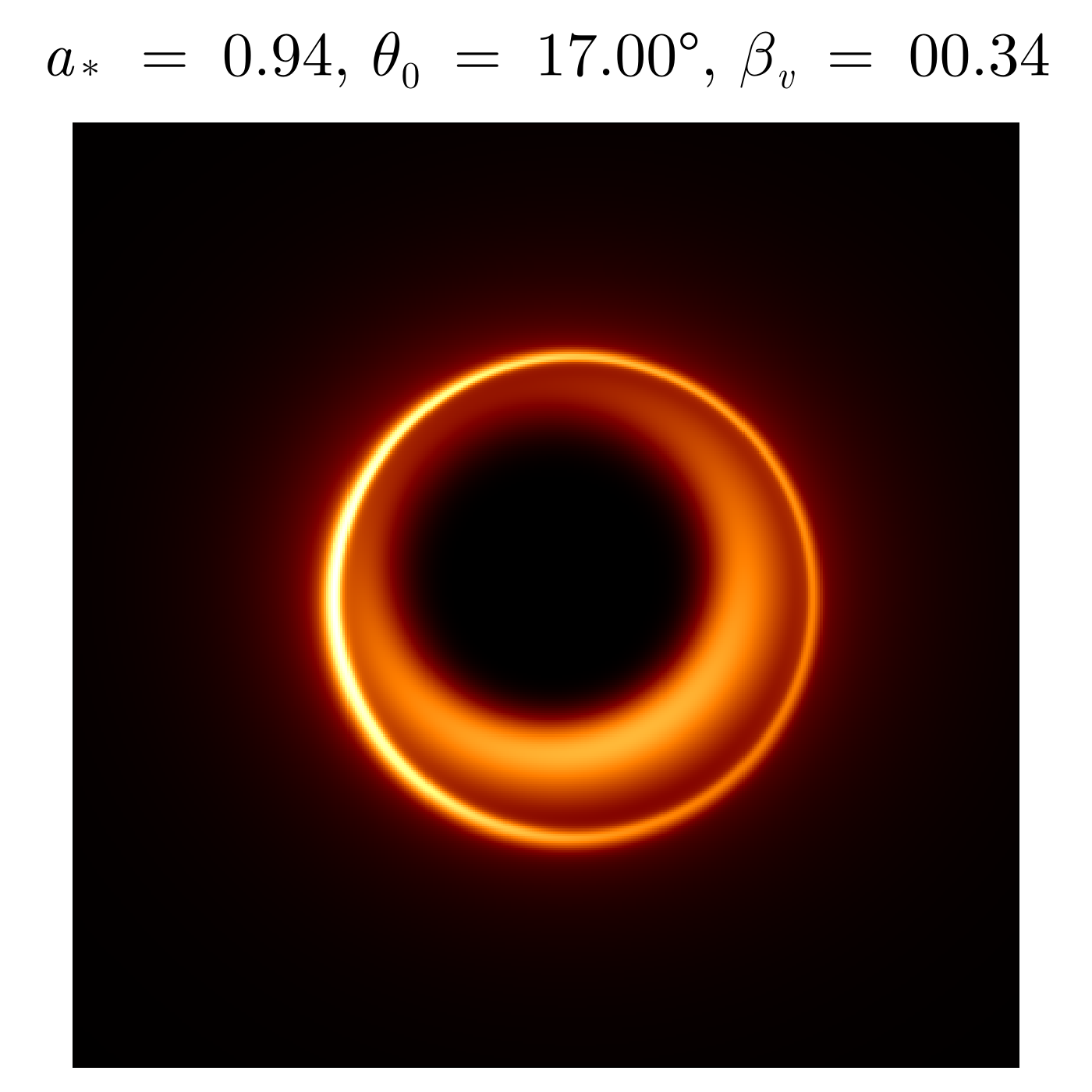}}
    \subfigure{\includegraphics[width=0.24\textwidth]{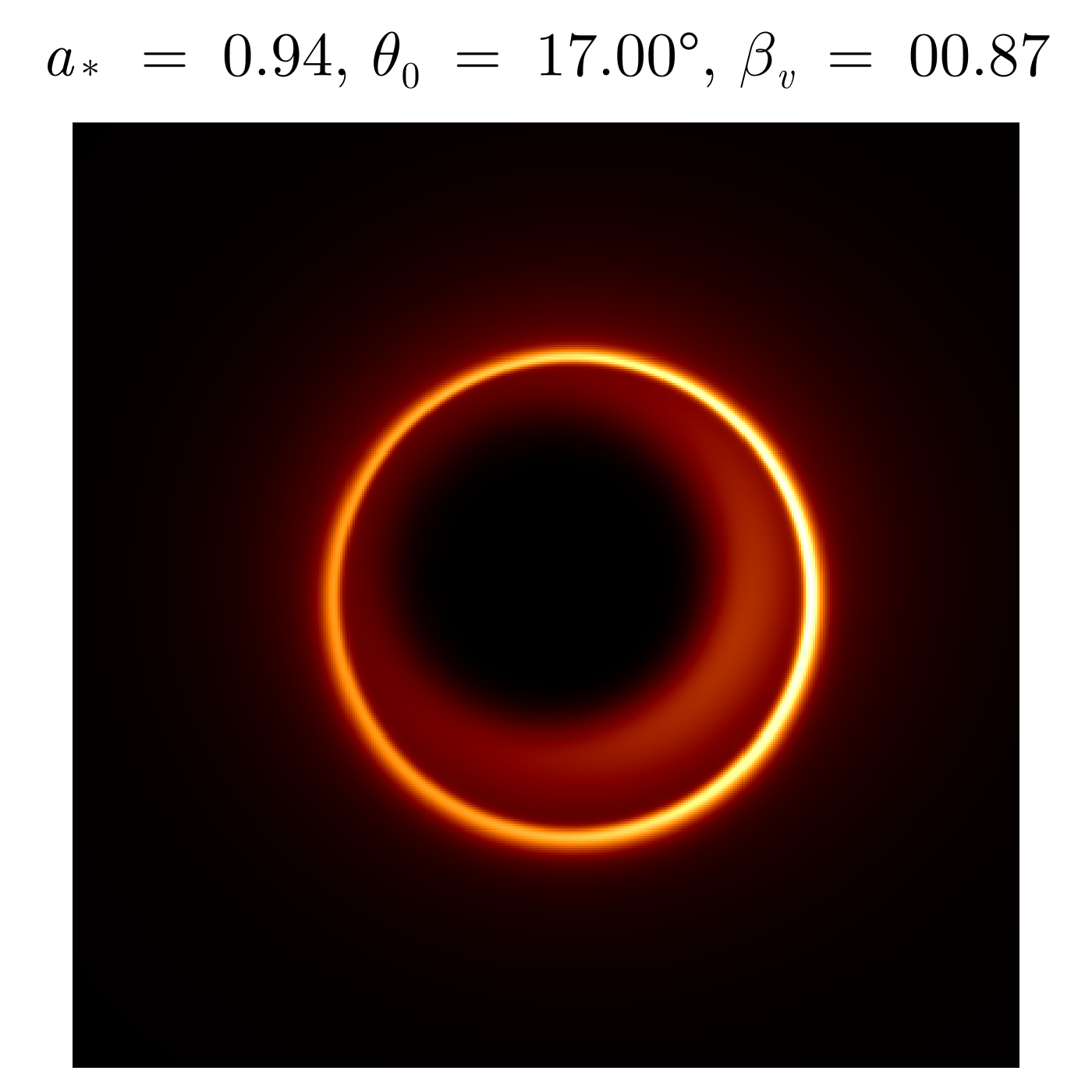}}
    \subfigure{\includegraphics[width=0.24\textwidth]{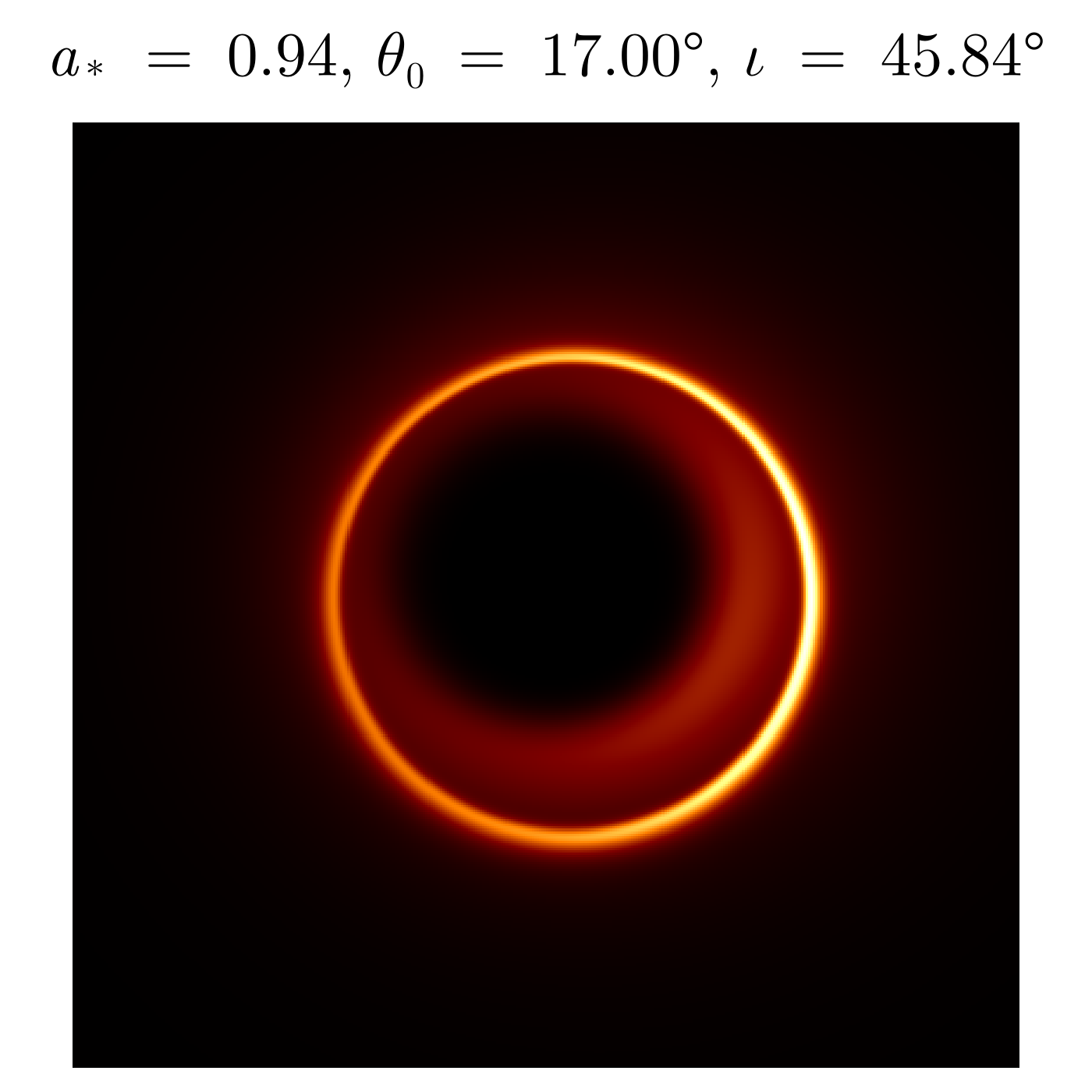}}
    \subfigure{\includegraphics[width=0.24\textwidth]{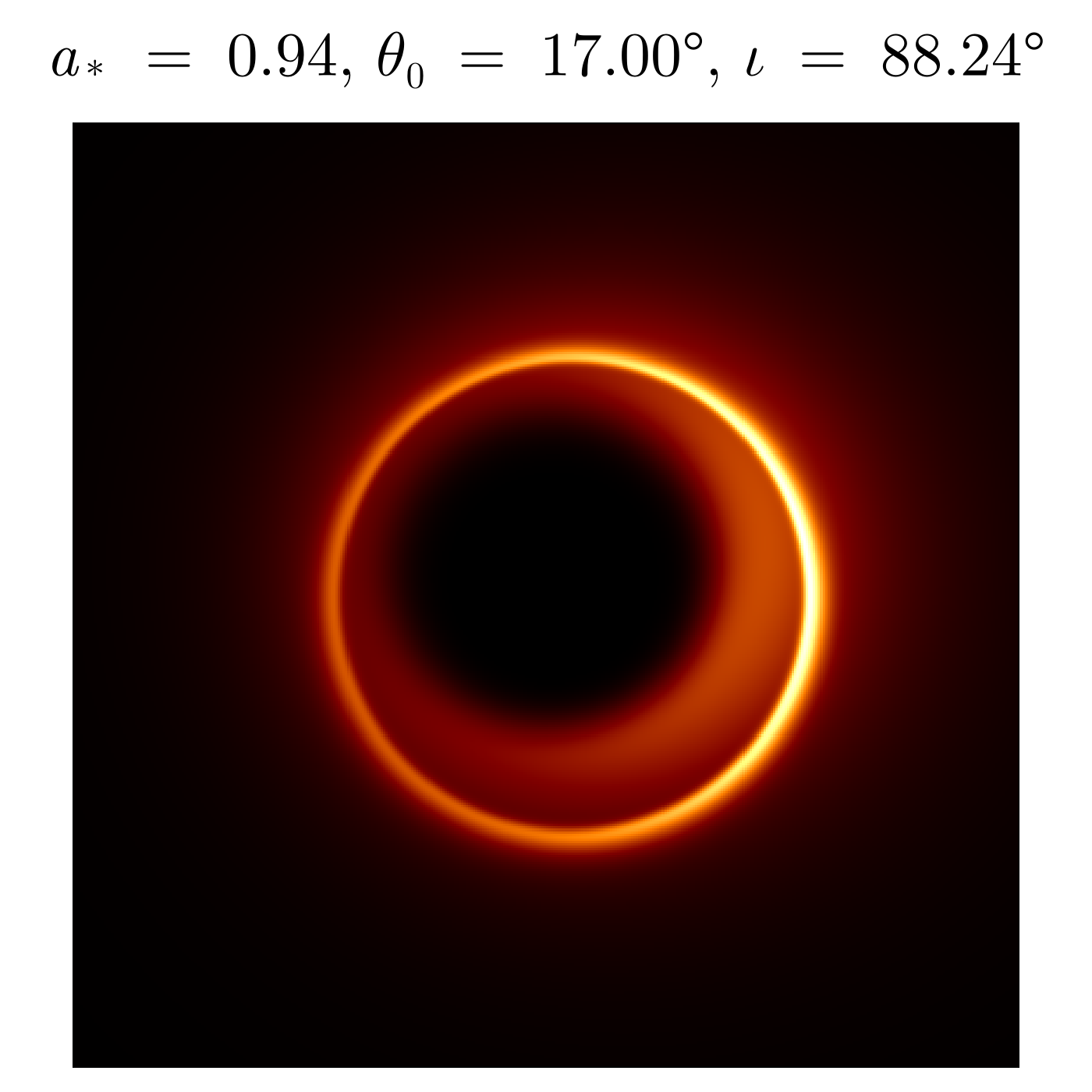}}
    \subfigure{\includegraphics[width=0.24\textwidth]{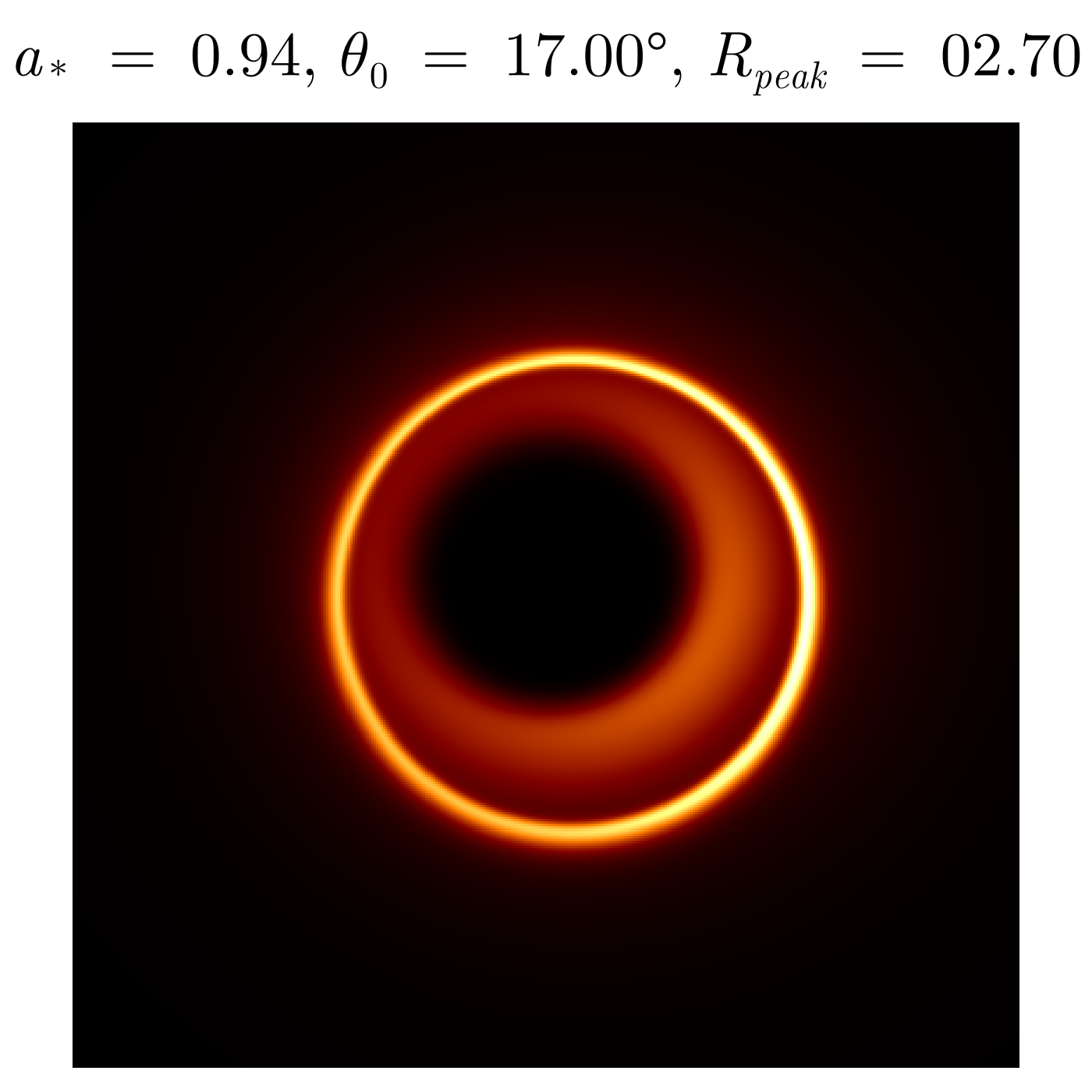}}
    \subfigure{\includegraphics[width=0.24\textwidth]{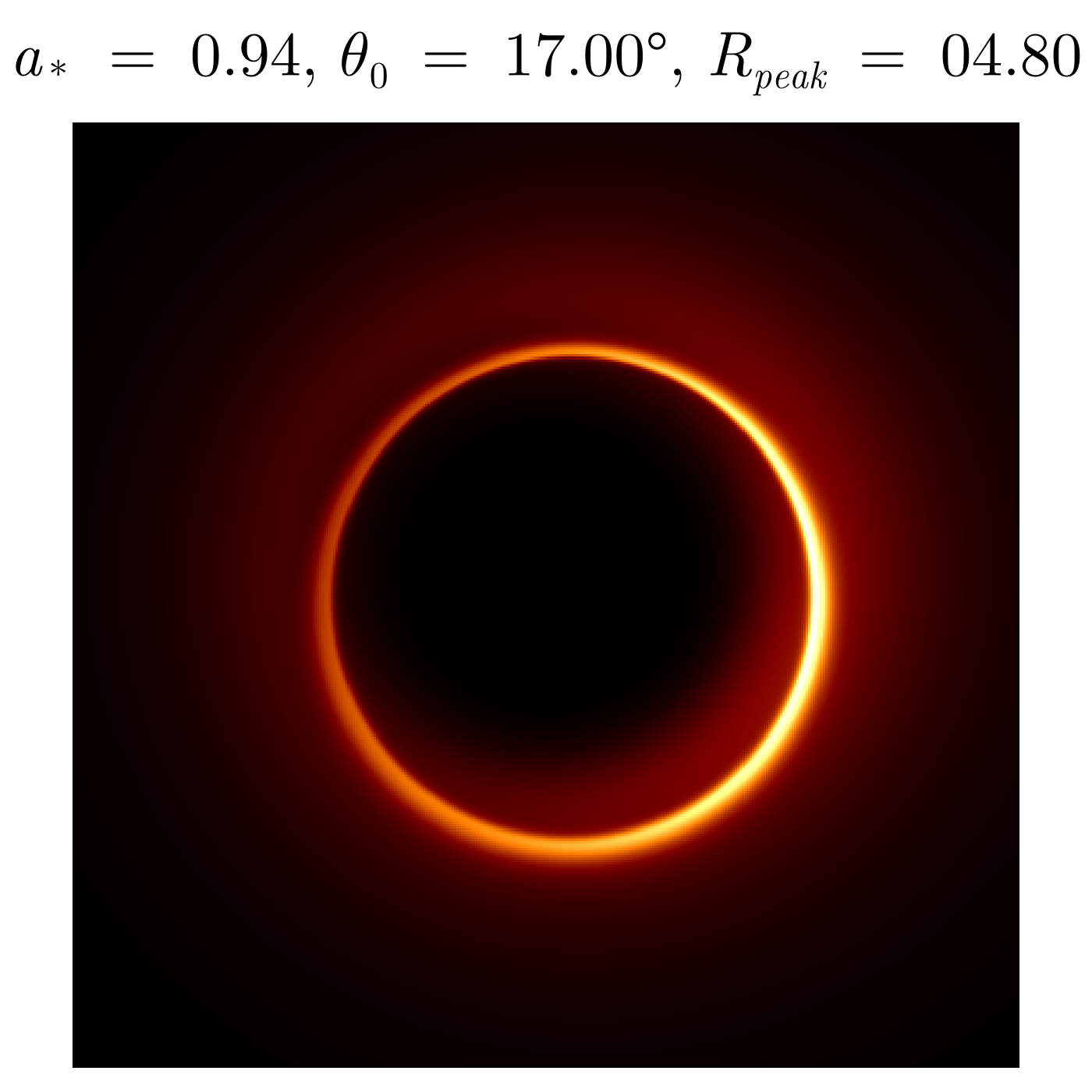}}
    \subfigure{\includegraphics[width=0.24\textwidth]{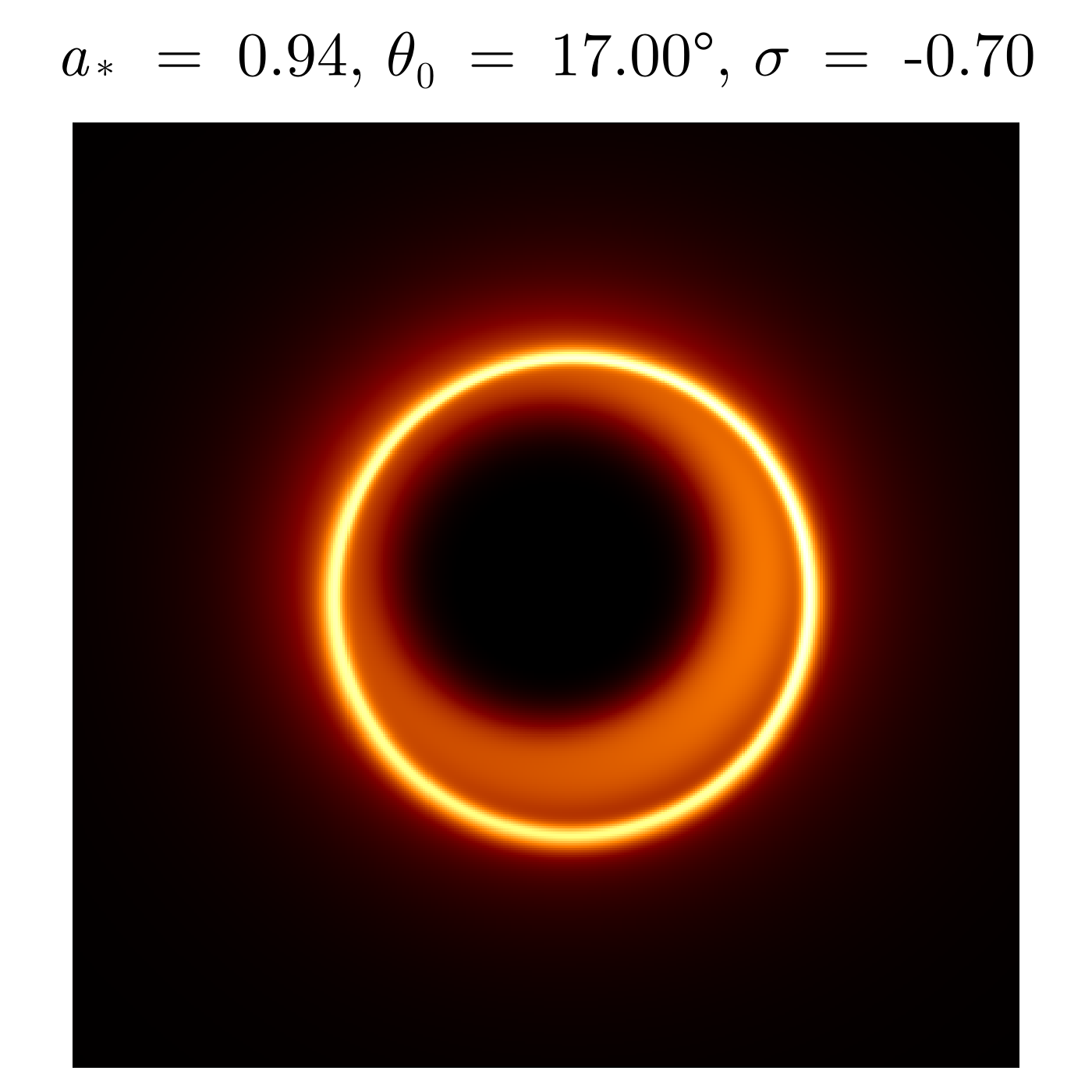}}
    \subfigure{\includegraphics[width=0.24\textwidth]{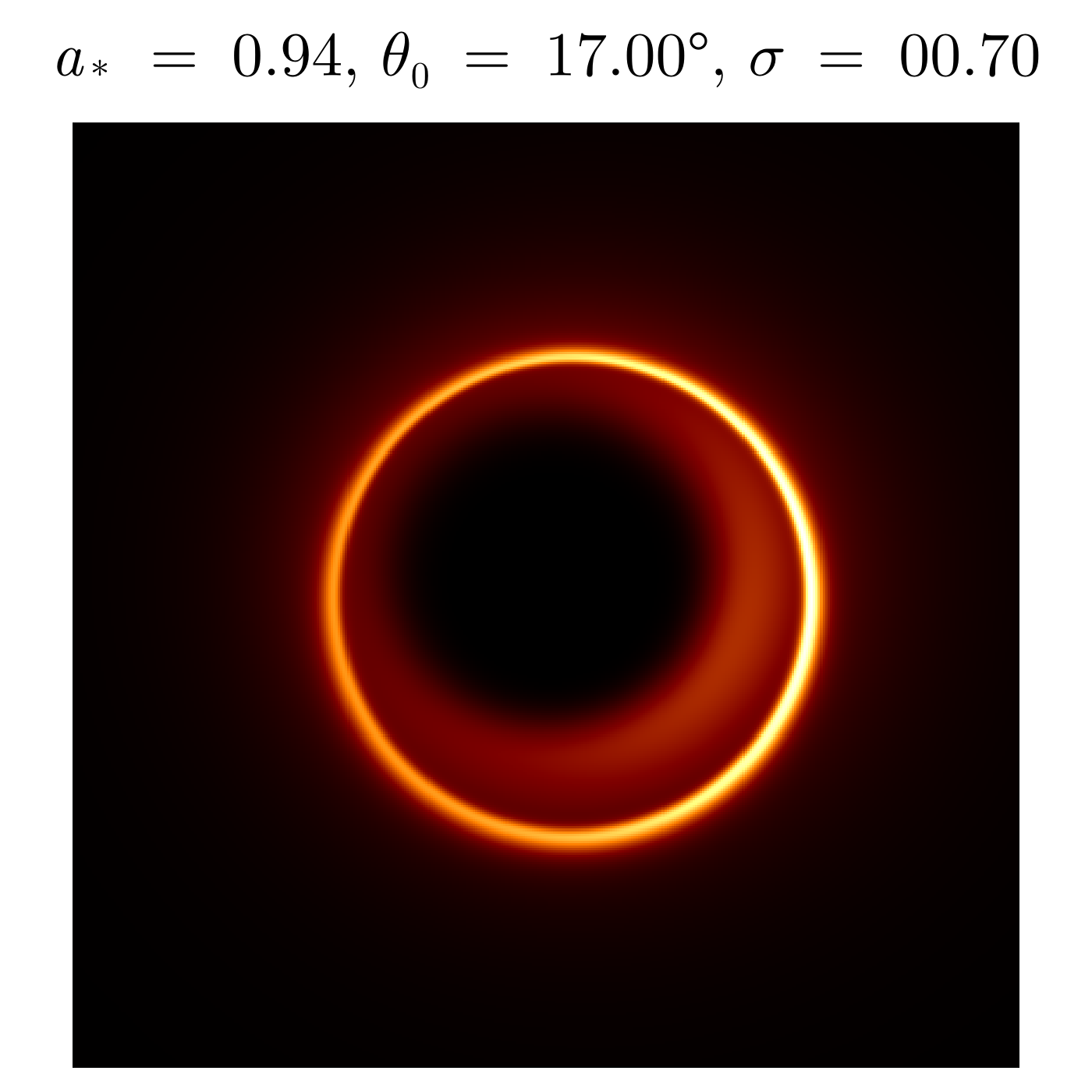}}
    \caption{Stokes I images, including both direct $n=0$ and indirect $n=1$ sub images, produced using the dual-cone accretion flow model. The spin, inclination, cone opening angle, azimuthal fluid angle, fluid speed, magnetic field orthogonality angle, characteristic emission radius, and spectral index are varied, forming a variety of image morphologies. While one parameter is varied, the rest remain fixed. In most images, there is a distinct brightness depression, or \textit{shadow}, and a thin, bright \textit{photon ring} feature.}
    \label{fig:krang-vary}
\end{figure*}

For our analysis, we produce images with both the direct emission ($n=0$ sub image) and the most direct photon ring emission ($n=1$ sub image). \autoref{fig:krang-vary} shows a sample of possible image morphologies resulting from different parameter combinations in the dual-cone accretion and emission model.
There is typically brightness asymmetry in both sub images. The centroid of the photon ring may be offset from that of the shadow. For \sgra, infalling matter emits synchrotron radiation up until very near the event horizon, where a noticeable brightness depression, or \textit{shadow}, is present \citep{Broderick_2016, Pu_2018, EHT2019-V}. Emission from \sgra is found to be largely consistent with models using Radiatively Inefficient Accretion Flows (RIAFs)/Advection Dominated Accretion Flows (ADAFs) \citep{2018ApJ...863..148P}. As such, the disk---unable to cool efficiently---heats to high temperatures ($\sim 10^{10}-10^{11}$ K) and puffs up, sometimes significantly. Due to the disk height, sub-Keplerian orbits are supported in the accretion flow, establishing inward radial momentum akin to a funnel. With RIAF models, we therefore observe $n = 0$ emission very near the black hole event horizon, not terminating at the ISCO as would be expected for other models, like a pressure-supported, radiatively efficient thin disk \citepalias{EHT2022-V, EHT2024-VIII}. The ``inner shadow'' present in intensity images is nearly indicative of the edge of the event horizon. We are able to observe this inner shadow because emission is largely confined to the equatorial plane with some height allowed in the dual-cone model; there is little obfuscation by the mostly evacuated region outside of the equatorial disk. With thin emission surfaces, the dual-cone model is capable of producing images consistent with RIAFs in a lightweight fashion.

Synchrotron emissivity is approximately strongest where the cross product $I \propto |\mathbf{k} \times \mathbf{B}|$ is maximized. With some parameter configurations, near-horizon emission is favored, since photons are mostly emitted with momentum aligned tangentially with the cones. As the dimensionless spin parameter $a_{\rm *}$ is increased in magnitude, the $n = 1$ sub image shrinks; the photon ring and ISCO move closer to the event horizon as an expected consequence. There is a translation of the photon ring with respect to the shadow, and brightness asymmetry profiles of both sub images vary with spin. There is a marked increase in the edge brightness of the $n = 1$ sub image and a slight increase in the $n = 0$ sub image brightness. Doppler beaming plays a significant factor in brightness distributions across both sub images. The $n = 0$ direct image is not significantly impacted by changing spin. The $n = 1$ sub image, however, where photons wrap around the black hole, is strongly dependent on the spin and the spacetime configuration around the black hole. The photon ring acts as a probe of null geodesics in a Kerr spacetime.

Image structure is also strongly dependent on inclination angle. For instance, an observer nearly face on with the spin axis sees little brightness asymmetry. Spin has a greater effect on brightness asymmetry at larger inclination angles (further from face-on). As the observer approaches an edge-on view, $n = 0$ direct emission becomes less prominent and the photon ring is translated downward and rightward compared to the inner shadow center. In our analysis, we constrain inclination angles to $\theta_o \in [1\degree, 89\degree]$, since larger angles can be produced with negative spin. 

All parameters form a subtle interplay of effects on ray-traced intensity images. For instance, models with opening angles $\theta_{\rm s}$ closer to equatorial have larger shadows and more dispersed $n=0$ direct emission. The photon ring is largely unaffected by alterations in cone opening angle.  The azimuthal direction of fluid flow $\chi$ has significant effects on the ray-traced image. With bulk fluid motion radially outward, there is a wide and bright contribution from direct emission. This contribution dims for retrograde flow, and is weakest for radially inward flows. Altering $\chi$ also affects brightness asymmetries and distributions in the $n=1$ sub image. As the magnetic field approaches equatorial (increasing $\iota$) and becomes more aligned with the emission source, there is a reduction in brightness of both the $n=0$ and $n=1$ sub images, especially for near-horizon emission. This is due to a reduction in the cross product $I \propto |\mathbf{k} \times \mathbf{B}|$. In certain parameter configurations, increasing $\beta_{\rm v}$ reduces the direct emission contribution and amplifies edge-brightening in the photon ring. Though not a direct measure of ring width, the outer index of the emissivity profile, $p_2$, strongly impacts the width of ring emission, which is historically more strongly constrained by the EHT \citepalias{EHT2022-IV}.

\section{Expanded Results}\label{sec:results-appendix}
\subsection{Posteriors from a Single Snapshot}\label{subsec:results-appendix-snapshot}

\begin{figure*}[!ht]
    \centering
    \includegraphics[width=\textwidth]{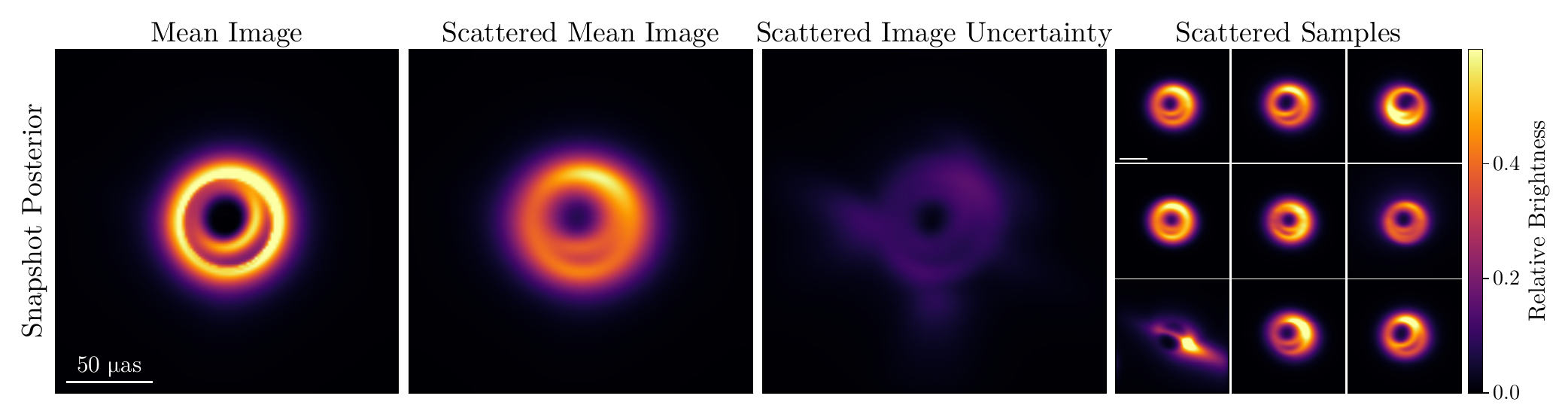}
    \caption{Example \sgra snapshot posterior results from a 120\,s segment of data starting at 13:49 UTC on April 7, 2017. Left to right: the mean image, mean image blurred with the diffractive scattering kernel, scattered image uncertainty, and scattered image samples drawn from snapshot posterior. The mean image includes expected features like an inner shadow, diffuse direct emission, and a bright photon ring. The photon ring is slightly offset from the centroid of the shadow. The standard deviation image and samples display the wide variation in possible image morphologies which fit the snapshot data. Images produced by the ray-traced accretion model have no fixed flux. We  unit normalize the flux for analysis, so the absolute brightness values shown are relative/nonphysical.}
    \label{fig:scan157-mean-std-img}
\end{figure*}

\begin{figure*}[ht!]
\centering
\includegraphics[width=0.95\linewidth]{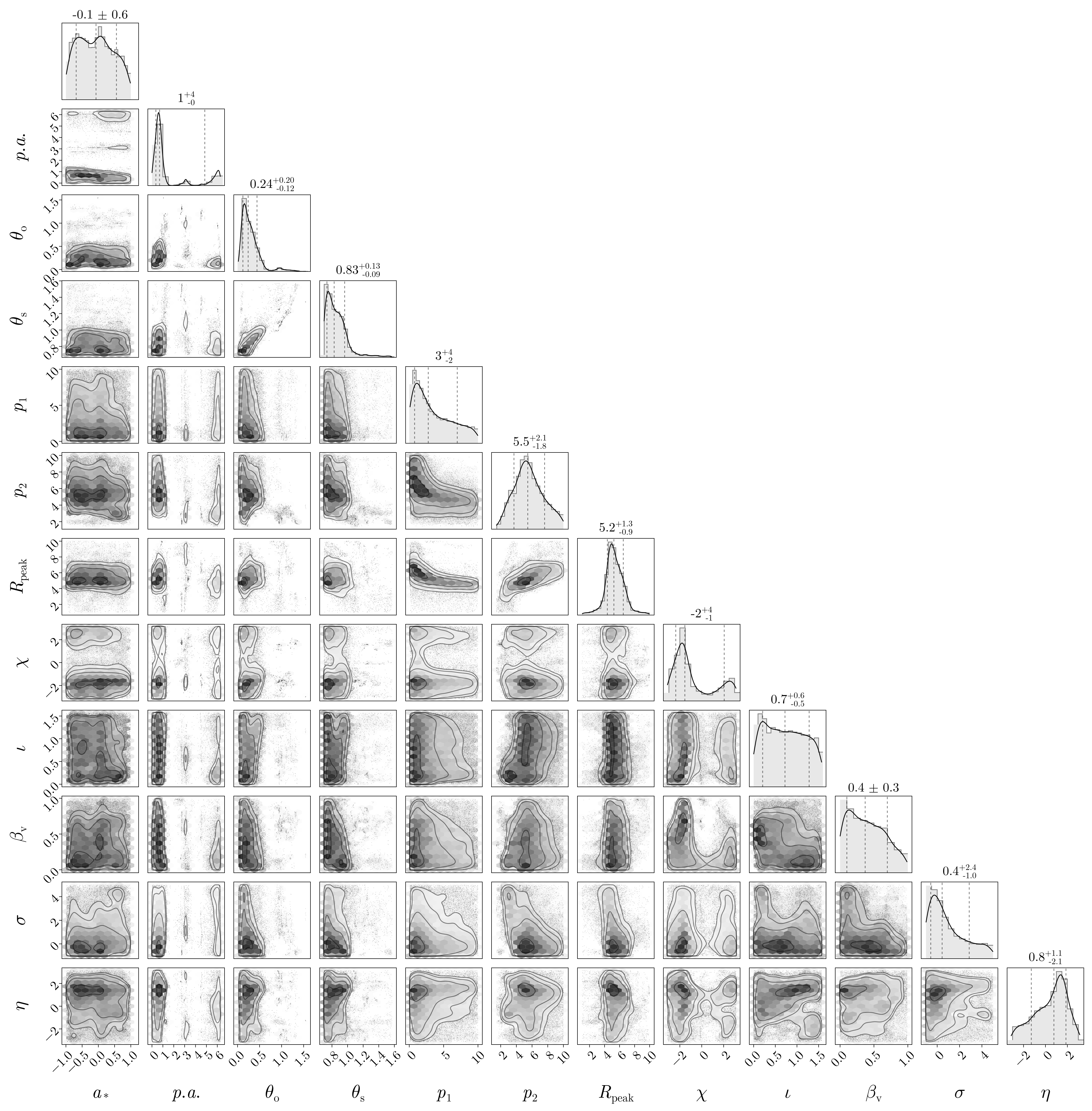}
\caption{Full pair plot for a single snapshot---scan 157 (April 7, 2017 at 13:49 UTC)---with good coverage (9 baselines). Spin is unconstrained. The position angle is well constrained at around $0$ or $2\pi$ with an additional mode near $\pi$. A nearly face-on inclination is favored in this snapshot, as well as a low cone opening angle. $p_1$ and $p_2$ are somewhat constrained ($p_2$ to a greater extent) with support across the entire prior. Fixing the mass-to-distance ratio resulted in a tightly constrained $R_{\rm peak}$ and eliminated multimodality in the $R_{\rm peak}$ posterior. $\chi$ is multimodal and largely unconstrained. $\iota$ and $\beta_{\rm v}$ are largely unconstrained, though $\beta_{\rm v}$ favors lower speed fluid flows. The spectral index and $\eta$ are also partially constrained in this snapshot,
\label{fig:scan157-full-pairplot}}
\end{figure*}

\autoref{fig:method-schematic} shows several images drawn from the snapshot posteriors of varying coverage. A large variety of image morphologies fit the data from a single snapshot, indicating the lack of parameter constrains from sparse visibility data. The various morphologies contribute to structures present in the mean and standard deviation images of \autoref{fig:scan157-mean-std-img}. The full corner pair plot from fitting snapshot scan 157 (April 7, 2017 at 13:49 UTC) is shown in \autoref{fig:scan157-full-pairplot}. While parameters such as the spin position angle, inclination, opening angle, outer index of the synchrotron emissivity profile, characteristic radius of the emissivity profile, and spectral index are relatively well constrained, others are poorly constrained. Most parameter posteriors show support across the prior range from a single snapshot fit. Despite the sparsity of data, low residuals between the actual visibilities and those reproduced by inference suggest our sampling stably and accurately analyzes the data.

\subsection{Hierarchical Model Posteriors}\label{subsec:hierarchical-posteriors-appendix}

\begin{figure*}[ht!]
\centering
\includegraphics[width=0.95\linewidth]{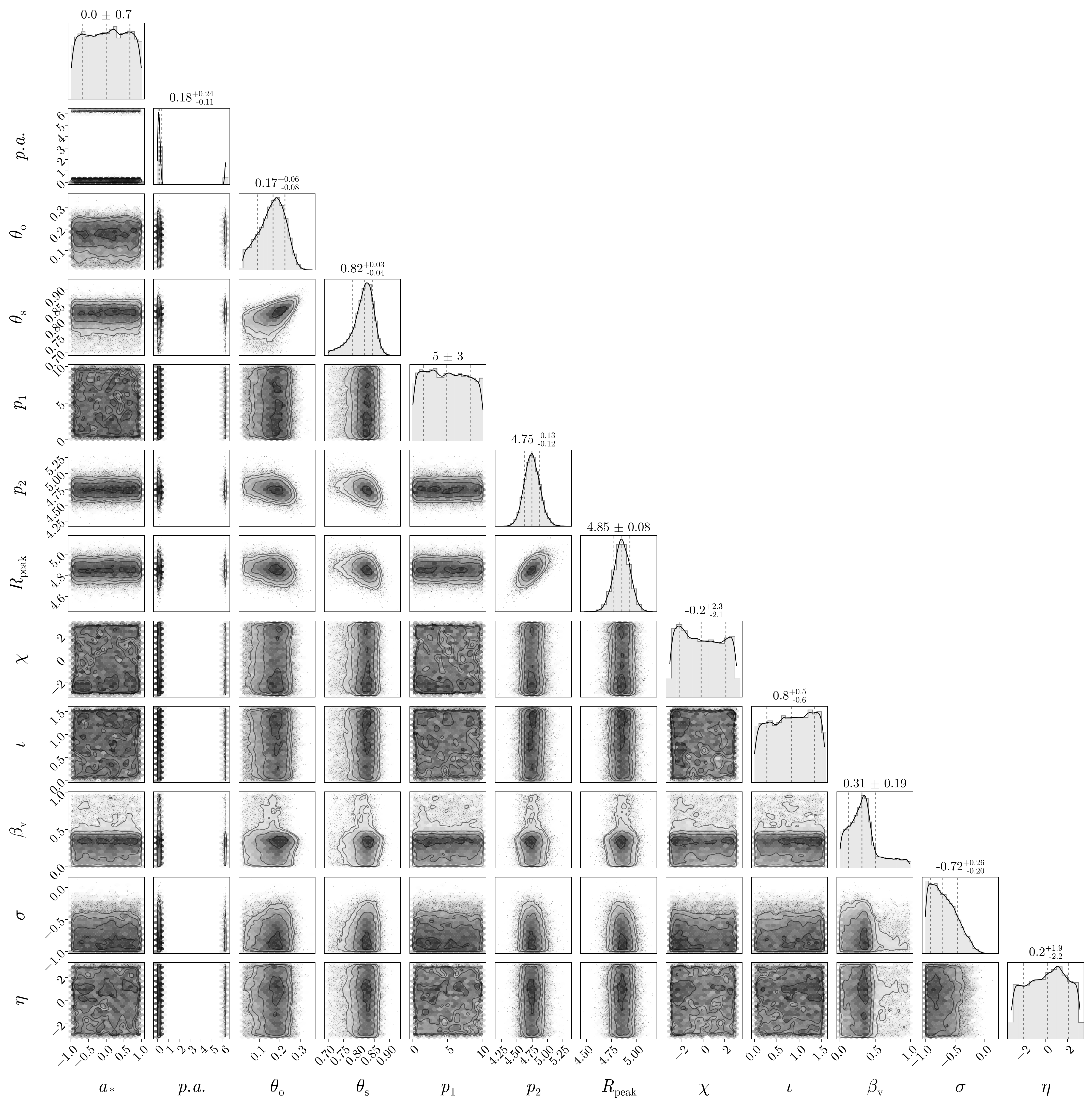}
\caption{Full pair plot for the global hierarchical model. Spin is entirely unconstrained. The spin position angle is highly constrained at $0$ or $2\pi$. A nearly face-on inclination and low opening angle are favored. Despite partial constraints in some snapshots, $p_1$ is largely unconstrained in the stacked model due to significant time variability. $p_2$---indicative of the ring width---and the characteristic emission radius are well constrained. There appears to be slight positive correlation between $p_2$ and $R_{\rm peak}$. Magnetic field parameters $\iota$ and $\eta$ are unconstrained in the global model. A low fluid speed $\beta_{\rm v}$ is preferred and a negative spectral index.
\label{fig:global-full-pairplot}}
\end{figure*}

The full pair plot for the average source structure distribution of the Bayesian hierarchical model is given in \autoref{fig:global-full-pairplot}. The posteriors provide parameter estimates and uncertainties for the persistent, time-averaged source structure. We obtain reasonable estimates and robust uncertainties for a variety of physical parameters.

\section{Inference on Synthetic GRMHD Data}\label{sec:grmhd-appendix}

\begin{figure*}[ht!]
\centering
\includegraphics[width=0.95\linewidth]{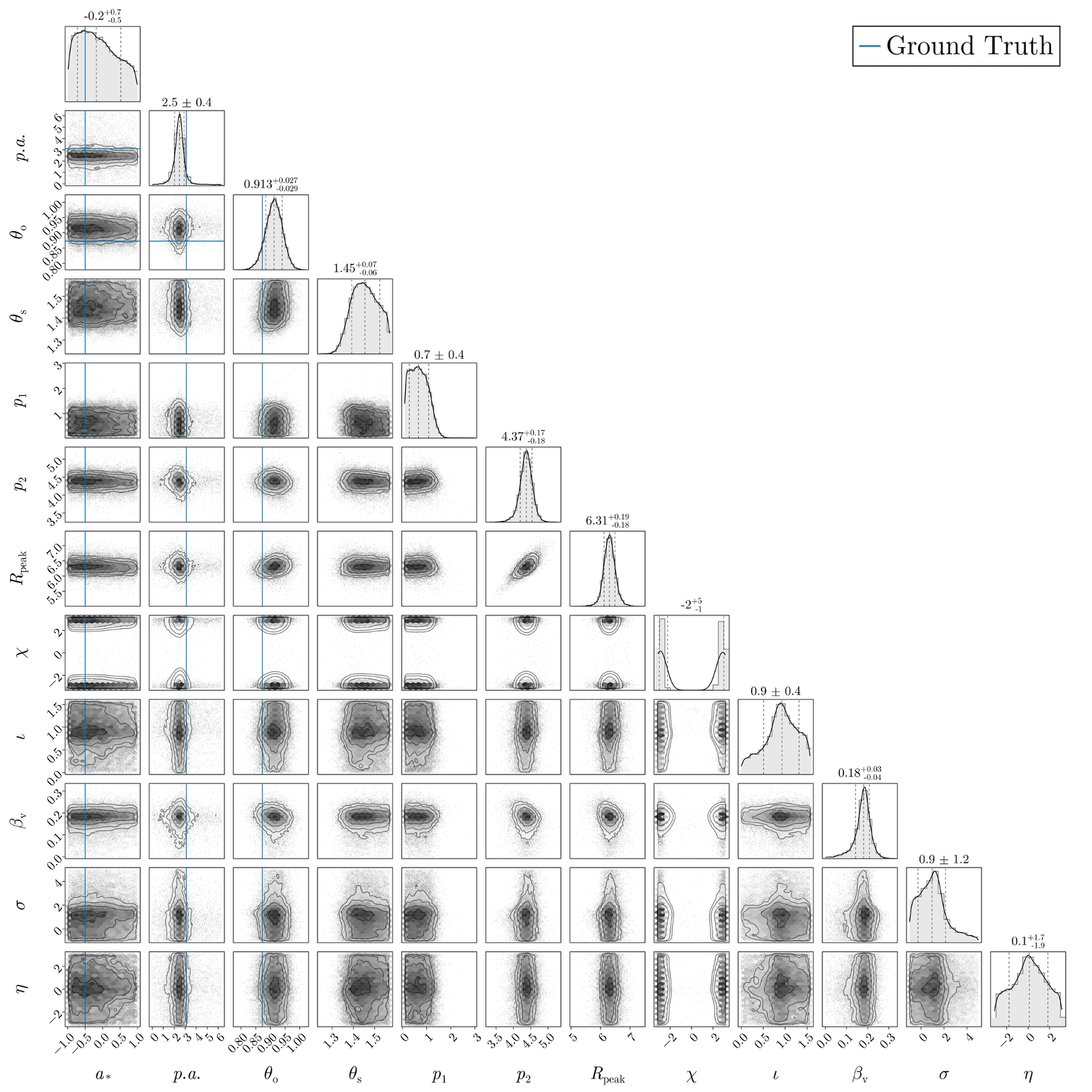}
\caption{Full pair plot for inference on synthetic visibilities generated from a time-variable GRMHD simulation. For parameters with known truth (black hole spin, spin position angle, and observer inclination), inference either does not constrain (in the case of spin) or nearly recovers the ground truth, revealing no significant biases in the inference pipeline for these parameters.
\label{fig:grmhd-full-pairplot}}
\end{figure*}

The full pair plot for inference on synthetic GRMHD data discussed in Section \ref{subsec:grmhd-inference} is shown in \autoref{fig:grmhd-full-pairplot}. The spin is unconstrained, with the known truth ($a_*=-0.5$) within the posterior support. The true spin position angle is nearly recovered with large uncertainty and a wide posterior. The observer inclination is tightly constrained and the true inclination is recovered well. The inferred cone opening angle indicates near-equatorial emission dominates the source, consistent with the MAD GRMHD simulation use to produce synthetic visibility data. Radially inward flows are strongly supported based on the azimuthal fluid angle, $\chi$, posterior. Interestingly, magnetic field parameters are more tightly constrained than in the real \sgra data analysis.

\bibliography{refs}{}
\bibliographystyle{aasjournalv7}

\end{document}